\documentclass[twocolumn]{aastex631}
\usepackage{makecell}
\usepackage{amsmath}

\begin{document}

\title{Little Red Dots and their Progenitors from Direct Collapse Black Holes}

\author[0000-0002-6038-5016]{Junehyoung Jeon}
\affiliation{Department of Astronomy, University of Texas, Austin, TX 78712, USA}
\author[0000-0002-4966-7450]{Boyuan Liu} 
\affiliation{Universit\"at Heidelberg, Zentrum fur Astronomie, Institut f\"ur Theoretische Astrophysik, D-69120 Heidelberg, Germany}
\author[0000-0003-0212-2979]{Volker Bromm}
\affiliation{Department of Astronomy, University of Texas, Austin, TX 78712, USA}
\affiliation{Weinberg Institute for Theoretical Physics, University of Texas, Austin, TX 78712, USA}
\author[0000-0001-7201-5066]{Seiji Fujimoto}
\affiliation{David A. Dunlap Department of Astronomy and Astrophysics, University of Toronto, 50 St. George Street, Toronto, Ontario, M5S 3H4, Canada}
\affiliation{Dunlap Institute for Astronomy and Astrophysics, 50 St. George Street, Toronto, Ontario, M5S 3H4, Canada}
\author[0000-0003-1282-7454]{Anthony J. Taylor}
\affiliation{Department of Astronomy, University of Texas, Austin, TX 78712, USA}
\author[0000-0002-5588-9156]{Vasily Kokorev}
\affiliation{Department of Astronomy, University of Texas, Austin, TX 78712, USA}
\author[0000-0003-2366-8858]{Rebecca L. Larson}
\affiliation{Space Telescope Science Institute, 3700 San Martin Drive, Baltimore, MD 21218, USA}
\author[0000-0002-0302-2577]{John Chisholm}
\affiliation{Department of Astronomy, University of Texas, Austin, TX 78712, USA}
\author[0000-0001-8519-1130]{Steven L.~Finkelstein}
\affiliation{Department of Astronomy, University of Texas, Austin, TX 78712, USA}
\author[0000-0002-8360-3880]{Dale D. Kocevski}
\affiliation{Department of Physics and Astronomy, Colby College, Waterville, ME 04901, USA}

\email{junehyoungjeon@utexas.edu}

\begin{abstract}
The \textit{James Webb Space Telescope (JWST)} has discovered a new population of objects, the Little Red Dots (LRDs), characterized by V-shaped spectra indicative of strong breaks around the Balmer limit and compact morphology that gave them their name. A popular explanation is that they are a sub-population of active galactic nuclei/supermassive black holes (AGN/SMBHs) predominantly found in the high-redshift Universe ($z\gtrsim3$). Similarly, direct collapse black holes (DCBHs), theorized to form from collapsing massive, extremely metal-poor gas clouds, have been invoked to explain high-redshift quasars, the most massive AGN sub-population. Here, we employ the semi-analytical code A-SLOTH to produce a population of DCBHs and compare them against observed LRD demographics and properties. Specifically, we compare the DCBH-seeded SMBH population against the standard stellar-remnant seeds and find that DCBH models agree better with observed LRD population statistics and host halo properties. Furthermore, for the most extreme and earliest LRD detections, interpreted to be systems with an AGN but little stellar component, DCBHs are able to reproduce the observed spectral shape and properties under multiple scenarios - high dust attenuation or AGN surrounded by dense gas - that have been proposed to explain the unique shape of LRD spectra. Even when super-Eddington accretion, invoked previously to explain the nature of LRDs, is enforced on stellar remnant seeds, the spectral characteristics of extreme LRDs cannot be reproduced. We emphasize the importance of gas-metallicity observations as an additional dimension besides the widely used SMBH-stellar mass ratios to further constrain the progenitors of LRDs.
\end{abstract}


\section{Introduction} \label{sec:intro}

The supermassive black holes (SMBHs) observed throughout the Universe must have formed earlier in cosmic history. With masses up to $\gtrsim10^8$ M$_\odot$ in the local Universe, these extreme objects most likely did not form initially with their immense masses, but instead formed as lower-mass black hole (BH) seeds and accreted baryonic material from the nearby environment, growing in mass and producing large amounts of radiation as active galactic nuclei \citep[AGN;][]{Soltan1982,Heckman2014,Hickox2018}. More specifically, observations show that most galaxies in the local Universe host such SMBHs at their centers \citep[e.g.,][]{Kormendy2013}. Then, to better understand how the SMBHs with their extreme properties and the galaxies that host them came to be, observations at early times/high redshifts are crucial, and with the launch of the {\it James Webb Space Telescope (JWST)} a multitude of new AGN at $z\sim5-10$ have been discovered \citep[e.g.,][]{Kocevski2023,Kocevski2025,Ding2022,Larson2023,Matthee2023,Onoue2023,Furtak2023,Bogdan2023,Juodbalis2023,Bosman2023,Greene2023,Kokorev2023,Kokorev2024,Fujimoto2023,Maiolino2023,Taylor2024,Chisolm2024,Naidu2025}.

Among such high redshift AGN observations, an intriguing population of objects termed Little Red Dots (LRDs) has been discovered. These ubiquitous and compact objects with multiple unambiguous permitted broad lines (which are very difficult to produce except from strong gas outflows or rotation caused by an AGN), visual morphology that match their name, and distinct V-shaped spectra \citep{Kocevski2023,Matthee2023,Labbe2024,Wang2024,Kokorev2024_lrd}, were not detected before \textit{JWST}. One possible explanation for their physical nature is that their rest-optical emission is powered by dust-obscured AGN \citep{Durodola2024,Huang2024,Kokorev2024}, although no dust continuum has been detected yet from LRD populations \citep[e.g.,][]{DeRossi_ALMA2023,Xiao2025, Setton2025, Casey2025,Akins2024}. Other scenarios, such as extremely dense stellar clusters, have also been proposed to explain LRDs \citep{Leung2024,Guia2024,Perez2024,Baggen2024}, but it is not clear whether such extremely compact stellar configurations could be stable. Surprisingly, the vast majority of newly discovered high-$z$ AGN are X-ray weak \citep{Kocevski2023,Matthee2023,Akins2024}, with typically only upper limits established so far, reminiscent of the Compton-thick AGN \citep{Fujimoto2022}. Although there are a few possible exceptions \citep[e.g.,][]{Kocevski2025}, this situation is in stark contrast to lower redshift typical AGN and quasars \citep[e.g.,][]{King2024,Pacucci2024}. 

Within the LRDs, there are select cases that are particularly puzzling, such as the ``BH star"  MoM-BH*-1 object \citep{Naidu2025}, showing the remarkable Balmer break strength of $\sim$2 AB magnitudes, which cannot be reproduced by any stellar populations, and is thus thought to be a system with just the central SMBH surrounded by dense, dust-free gas. The extremely dense gas is able to imprint signatures in the observed spectrum similar to that of a stellar population \citep{Inayoshi2025_bl}. Furthermore, the extremely dense gas may cause electron scattering, which is also nicely aligned with the exponential line profile observed in the deep spectra taken for LRDs \citep{Rusakov2025}\footnote{The exponential profile results in lower full width at half-maximum of broad lines, which may indicate that the LRD BH masses have been overestimated.}. 
A lower redshift LRD named ``The Cliff" has also been observed with similarly remarkable Balmer break stength as MoM-BH* \citep{deGraff2025}, and such an unusual LRD has now been reported to exist even out to $z=9.3$, CAPERS-LRD-z9 \citep{Taylor2025}. It remains mysterious how such a massive BH of $\sim10^{7.5}$ M$_\odot$, assuming the local scaling of broad-line widths to BH mass from \citet{Greene2005}, could have formed so early in cosmic history.

In this work, we explore whether heavy direct collapse black holes (DCBHs) can be the progenitors of systems like the peculiar MoM-BH*-1 and CAPERS-LRD-z9 objects, and in turn a subset of the LRDs. DCBHs have first been proposed to explain the massive SMBHs inferred to power luminous quasars (log $M_{\rm BH}/M_{\odot} >$ 8) in the early Universe ($z\gtrsim 6$), already prior to \textit{JWST} observations \citep[e.g.,][]{Wu2015,Banados2018,Smith2019_2,Woods2019,Inayoshi2020,Zubovas2021,Fan2023}. They are predicted to form from the runaway collapse of a massive, extremely metal-poor \citep[$Z\lesssim 10^{-3}\ \rm Z_\odot$;][]{Chon2024} gas cloud, involving one or a few supermassive stars as an intermediate, short-lived stage \citep[e.g.,][]{Bromm2003,Begelman2006,Lodato2006}. This scenario, resulting in a larger BH seed mass ($\sim10^4-10^6$ M$_\odot$), relies on the rare conditions that suppress low-temperature gas cooling mechanisms, allowing the cloud to collapse without fragmenting to form a large number of ordinary/low-mass stars \citep[e.g.,][]{Lodato2007,Johnson2013,Wise2019,Haemmerl2018,Haemmerle2020,Luo2020}. 

As DCBHs start with a larger initial mass, they are subject to weaker timing constraints in growing to the observed high-$z$ SMBH masses \citep[e.g.,][]{Haiman2001,Johnson2007-2}. In addition, this seeding model could also naturally explain the inferred ``overmassive'' systems, where the BH to stellar mass ratio is much higher than in the local Universe, by forming a massive SMBH in an environment with initially few stars \citep{Durodola2024,Jeon2024}. Many \textit{JWST}-discovered AGN, including the LRDs, when considering their inferred BH and stellar masses, exhibit such overmassive configurations, subject to uncertainties in mass measurement methodology \citep[e.g.,][]{Bogdan2023,Kokorev2023,Pacucci2023,Natarajan2023,Taylor2025}. Similarly, objects like MoM-BH*-1, inferred to host no stars, but an AGN residing in dense gas, could be interpreted as a natural consequence of the DCBH formation scenario\footnote{Primordial black holes (PBHs), formed through collapsing overdensities soon after the Big Bang, could provide an alternative pathway for early massive SMBH formation \citep[see e.g.,][]{Dayal2024,Zhang2025,Qin2025}}. 

Therefore, we aim to compare DCBH models against existing LRD observations to assess whether DCBHs could be the progenitors of LRDs and systems like MoM-BH*-1. We specifically utilize the A-SLOTH (Ancient Stars and Local Observables by Tracing Halos) code \citep{Hartwig2022,Hartwig2024,Magg2022,Liu2025}, a semi-analytic model (SAM) designed to investigate the formation and evolution of the first stars, and particularly tuned to high-redshift constraints. We have developed SMBH/AGN formation and growth models, including DCBH seeds, to be used in the A-SLOTH framework to study the co-evolution of the SMBH, stellar, and halo populations, with a calibration approach to reproduce existing high-$z$ BH mass function (BHMF) observations \citep{Jeon2025,Taylor2024}.

This paper is organized as follows. In Section~\ref{sec:methods}, we introduce A-SLOTH and our custom-designed prescriptions to incorporate the DCBH physics into the SAM. We compare our model predictions with various LRD observations in Section~\ref{sec:results}, such as their overall demographics and spectral properties. In Section~\ref{sec:discussion}, we broadly assess the plausibility of explaining LRDs with heavy DCBH seeds. We summarize and conclude in Section~\ref{sec:conclusions}.

\section{Methodology} \label{sec:methods}

We use the semi-analytical A-SLOTH framework to explore the formation and evolution of a heavy DCBH seed population.  A-SLOTH is designed to model high-redshift galaxy formation, based on halo merger trees from N-body simulations or the Extended Press-Schechter (EPS) formalism \citep{Parkinson2008}. The code has been calibrated to reproduce the cosmic star formation rate density at $z\sim 4.5-13.3$, the optical depth to Thomson scattering, and properties of the Milky Way \citep{Hartwig2024}, and is focused on the state-of-the-art modeling of the first stars, so that BH formation and evolution in early cosmic history can be more accurately followed. Here, we extend the investigation of early BH seeding and growth in \citet{Jeon2025}, calibrated to high-redshift BHMF observations \citep{Taylor2024}, briefly summarizing the relevant A-SLOTH modeling below. We refer the reader for full details to the public A-SLOTH release papers \citep{Hartwig2022,Magg2022}, and for the BH physics prescriptions to \citet{Jeon2025}. The models in this work are again calculated with the EPS approach, to be able to explore a comprehensive parameter space with a reasonable computational cost and time.

\subsection{Black hole seeding}\label{sec:seeding}

For our investigation, we consider two BH seed formation channels: light seeds originating from Population~III (Pop~III) stellar remnants ($\sim100$\,M$_\odot$) and heavy DCBH seeds from collapsing massive clouds, likely involving a supermassive star (SMS) progenitor ($\sim10^5$\,M$_\odot$). Other scenarios for heavy seed formation have been proposed, such as SMBH formation through runaway collisions of (proto-)stars or BHs in dense stellar clusters \citep[e.g.,][]{Zwick2023,Reinoso2023,Klessen2023,Gaete2024}. We do not consider this channel in this work as we focus on producing BH*-like sources, which are predicted to have little to no stars. 

Our specific process of assigning BH seeds to halos is as follows:
Each halo in the merger tree is checked to see if any of the halo's progenitors hosted a BH previously. If no progenitors hosted a BH or if the halo is the very first progenitor, we evaluate the halo properties and determine whether it should be seeded with a BH and if so what kind. When a massive ($>40\ \rm M_\odot$) Pop~III star in the halo dies and its mass $m_\star$ is outside the pair-instability supernova range (producing no remnants), such that it falls within the $40$ M$_\odot<m_\star<140$ M$_\odot$ or $m_\star>260$\,M$_\odot$ mass windows, a light seed of the same mass as the dying Pop~III star is assigned at the halo center. We do not consider seeds from Pop~II stars for simplicity, and as they will typically result in much lower-mass ($\sim5-10$ M$_\odot$) BHs \citep{Stacy2016,Volonteri2021,Latif2022,Sassano2021}. We adopt the power-law Pop~III initial mass function adopted in A-SLOTH, as follows:
\begin{equation}
    \frac{dN}{d\log m_{\star}} \propto m^{-1}_{\star}\ ,\ m_\star\in[13.6,197]\ \rm M_\odot \mbox{\ .}
\end{equation}

For heavy seeds, we consider multiple criteria based on the halo virial temperature, metallicity, and Lyman-Werner (LW) feedback to capture the dense, hot, and metal-poor conditions required for DCBH formation, and ensure that the gas in the halo is not able to cool and fragment too quickly to form star clusters \citep{Bromm2003,Ardaneh2018,Wise2019,Chon2021}. Specifically, we require that the halo virial temperature is above the atomic cooling limit, $10^4$\,K, to be able to host gas that can collapse (nearly) isothermally even in the absence of H$_2$ cooling. We further require that the metallicity of the star-forming gas in the halo be smaller than the critical value of $Z<Z_{\rm crit} = 2\times10^{-4}~{\rm Z}_\odot$ \citep[e.g.,][]{Liu2020} to suppress efficient metal cooling. Finally, for the LW background, we require its flux to be greater than a critical level, $J_{\rm LW}>J_{\rm crit}$, so that LW radiation can dissociate H$_2$ even in dense, collapsing gas, thus disabling molecular cooling and preventing low-mass star formation. 

We set $J_{\rm crit}=300$, in units of $10^{-21}$ erg s$^{-1}$ cm$^{-2}$ Hz$^{-1}$ sr$^{-1}$, and we consider both global and local LW contributions. The global LW background is expressed as (in the same units) 
\begin{equation}\label{globallw}
    J_{\rm global} = 10^{2-z/5} \mbox{\ ,}
\end{equation}
and is generally subdominant to the local background, but we include it for completeness, based on \citet{Greif2006}. The local LW flux within a given halo provides the main contribution, and is calculated from considering stars above 5\,M$_\odot$ that are capable of producing LW radiation ($11.2-13.6$~eV) efficiently. We determine the LW photon production rate of each massive star based on the stellar mass from the fitting formula in \citet[][see their equ.~8]{Deng2024}, further assuming for simplicity that the high-mass stars are located on average at 0.1 times the virial radius away from the halo center. Thus, for a given halo, its LW flux is the sum of the global background and the local component, produced by the massive stars in the halo. Our model represents the optimistic DCBH seeding scenario where the local massive stars have formed in the progenitor halos, whose mergers heat the gas to $\sim 10^4$~K, triggering prompt DCBH formation within $\sim 1$~Myr in the post-merger halo, before the gas reservoir is destroyed by feedback (see \citealt{Sullivan2025} for a pessimistic scenario). 
If all the above criteria are met, regarding virial temperature, metallicity, and LW radiation, a heavy seed of mass $10^5$ M$_\odot$ forms at the halo center \citep[e.g.,][]{Becerra2018b,Becerra2018a}. The median cold gas mass in halos right after DCBH formation is $\sim5\times10^{4}$ M$_\odot$, of the same order as the initial DCBH mass. This agrees with the theoretical scenario that DCBH formation should take up most of the available cold gas in the host halo \citep{Wise2019}.

If any of the halo's progenitors contains a BH, the halo inherits at its center the BH from the most massive progenitor. If the most massive progenitor hosts multiple BHs, the other BHs are also inherited at their respective positions. If multiple progenitor halos host BHs, the BHs from the less-massive progenitors are all inherited as well, but placed at random (apocenter) distances from the halo center. In assigning distances, we follow the spatial distribution of Pop~III remnants derived from high-resolution simulations \citep{Liu2020,Liu2020_2} for lower mass BHs ($M_{\rm BH} < 10^5$\,M$_{\odot}$), and the locations of nuclear star clusters (NSCs), after halo mergers, found in previous A-SLOTH implementations \citep{Liu2024} for higher mass BHs ($M_{\rm BH} \geq 10^5$\,M$_{\odot}$). We have adopted the NSCs as tracers of post-merger massive BH locations, as NSCs are thought to reside in the satellite halo centers, similar to the massive BHs \citep{Partmann2024,Askar2024,Chen2024}. Finally, the BH orbits are assigned random eccentricities drawn from a uniform distribution in $[0,1)$ \citep{Liu2024}. Given the orbits of BHs in the (post-merger) halo, we follow their evolution by dynamical friction, as described in Sec.~\ref{sec:dynamics} below.

\subsection{Black hole accretion}\label{sec:accretion}

At each timestep, we update the BH mass and location through accretion and  dynamical friction. These steps are crucial to be able to model BH evolution, but we lack direct information on the gas distribution near the BH in SAMs. Therefore, we introduce some additional assumptions to determine the accretion rate. 

We first constrain the accretion rate to be limited by the available cold gas mass in the halo as
\begin{equation}\label{eq:macc}
    \delta M_{\rm BH} = \text{min}\left(f_{\rm duty}\dot{M}_{\rm acc}\delta t, M_{\rm cold}\right)\mbox{\ ,}
\end{equation}
where $M_{\rm cold}$ is the cold gas mass, $\delta t$ the timestep, and $f_{\rm duty}$ the duty cycle for \textit{maximal} active accretion onto the SMBH \citep{Pacucci2023,Lai2024}. In principle, the duty cycle is a free parameter, but has been calibrated in previous work to reproduce existing observations \citep[see][and Section~\ref{sec:discussion}]{Jeon2025}. 

To determine $\dot{M}_{\rm acc}$, we consider the fiducial limit of accretion, the Eddington rate, as a baseline to represent optimistic BH growth scenarios:
\begin{equation}
\dot{M}_{\rm Edd} = 2.7\times10^{-3}\left(\frac{M_{\rm BH}}{10^5~\text{M}_\odot}\right)\left(\frac{\epsilon_r}{0.1}\right)^{-1}\rm~M_\odot~\text{yr}^{-1}\mbox{\ .}
    \label{edd}
\end{equation}
We then parameterize the actual rate with the Eddington ratio $f_{\rm Edd}$ as follows:
\begin{equation}
    \dot M_{\rm acc} =  f_{\rm Edd}\dot{M}_{\rm Edd}
        \label{edd_acc}\mbox{\ .}
\end{equation}
Here, $\epsilon_r=0.1$ is the radiative efficiency, and we allow the Eddington ratio $f_{\rm Edd}$ to be larger than 1, corresponding to super-Eddington accretion. This parameter has been calibrated to be around 1.5 with $f_{\rm duty}=0.8$ in previous work to match the most extreme, highest redshift ($z\sim10$) AGN \citep{Jeon2025}. We note that evaluating the Eddington accretion rate does not require direct knowledge about the gaseous environment near the BH, and is mainly dependent on the current BH mass. We use this accretion implementation to model the most extreme SMBHs that the Universe may produce.

For a more physically detailed model, tied to the conditions in the surrounding gas, we use the Bondi-Hoyle formalism \citep{Bondi1944}:
\begin{equation}\label{bondi}
\dot{M}_{\rm Bondi} = \alpha\frac{4\pi(GM_{\rm BH})^2\rho_g}{c_s^3}\mbox{\ ,}
\end{equation}
\\
where $\rho_g$ is the gas density, $c_s$ the sound speed, and $\alpha$ the boost factor, which is a free parameter. The boost factor accounts for the enhanced gas density in the halo inner regions near the central BH that is not well captured in cosmological simulations \citep{Jeon2023}, or with idealized halo profile models \citep{Trinca2022}. We here set it to unity, given that A-SLOTH explicitly models the cold gas in the halo with near-CMB temperatures \citep{Safranek2016}. For $\rho_g$, we only consider the cold gas mass $M_{\rm cold}$ as contributing to BH accretion, and assume that it is confined to the halo scale-radius, $R_s = R_{\rm vir}/c_{\rm DM}$, of the Navarro-Frenk-White (NFW) dark matter halo profile \citep{Navarro1996}. Here, $c_{\rm DM}$ is the halo concentration parameter \citep{Hartwig2022}, given by the fitting functions from \citet{Correa2015}. We approximate the cold gas density distribution as an isothermal sphere with a flat core \citep{Trinca2022} 
\begin{equation}\label{isothermal}
    \rho(r) = \frac{\rho_{\rm 0}}{1+(r/r_{\rm core})^2}\mbox{\ ,}
\end{equation}
where $r_{\rm core} = 0.012R_{\rm vir}$ is the halo core radius and $\rho_{\rm 0}$ the normalization density. This density is set so that the integral of Equation~(\ref{isothermal}) up to the scale radius $R_s$ equals the total $M_{\rm cold}$. 

We evaluate this expression at the BH Bondi radius, $r_{\rm B} = GM_{\rm BH}/c_s^2$, where the sound speed is $c_s = \sqrt{k_{\rm B}T/\mu m_{\rm H}}$, with $k_{\rm B}$ being the Boltzmann constant, $T$ the gas temperature, $m_{\rm H}$ the atomic hydrogen mass, and $\mu$ the mean molecular weight of the gas. To approximate the gas temperature, we use the halo viral temperature plus an effective contribution expressing the heating from BH feedback, $T_{\rm vir} + \delta T_{\rm feed}$, when the average metallicity is below $Z_{\rm crit}$. If the latter is above $Z_{\rm crit}$, we instead employ the cold gas temperature at the given redshift plus the BH feedback contribution, $T_{\rm cold} + \delta T_{\rm feed}$, reflecting the fact that at high redshifts, the cold gas in the halo should be able to efficiently cool to temperatures lower than the halo virial temperature. The cold gas temperature $T_{\rm cold}$ is set to the cosmic microwave background (CMB) temperature \citep[e.g.,][]{Safranek2016}. Such cold gas near the CMB temperature has been found in high-resolution simulations even at lower redshifts of $z\sim5-6$ \citep{Liu2020,Jeon2023}. Invoking the thermodynamics of the CMB to impose a temperature floor \citep[see also][]{Johnson2006} provides a robust upper bound on accretion.

The additional heating from BH feedback, expressed in the equivalent $\delta  T_{\rm feed}$, is described in Section~\ref{sec:feedback}. We further limit the accretion rate to a multiple of the Eddington value (Equ.~\ref{edd}) as 
\begin{equation}
    \dot M_{\rm acc} = \text{min}\left(\dot{M}_{\rm Bondi}, f_{\rm Edd}\dot{M}_{\rm Edd}\right)\mbox {\ .}
    \label{bondi_acc}
\end{equation}
Using the Bondi-Hoyle equation has the advantage of adapting the BH growth to the physical conditions in its environment. However, unlike for the Eddington model, idealized estimates have to be used for the cold gas density and temperature, as such information is not directly available in SAMs. We calibrate $f_{\rm duty}=0.5$ for the Bondi-Hoyle accretion models \citep{Jeon2025}. 

With $\dot{M}_{\rm acc}$ being thus determined, an equivalent mass is removed from the cold gas reservoir of the halo. Accretion is only applied to the primary BH, which is assumed to reside in the halo center, where the dense and cold gas is also located. The other BHs are assumed to not accrete for simplicity, following previous work showing that wandering BHs that do not reside in the dense central region generally do not accrete efficiently and remain dormant \citep{Jeon2023,Ogata2024}. 

\subsection{Black hole dynamics and mergers} \label{sec:dynamics}

If a BH is not at the center of a halo after mergers, we follow its inspiral and update its location and eccentricity at each (star formation) timestep using the dynamical friction prescription from stars and DM in \citet[][equations 2-4]{Liu2024}. For a more complete description of the treatment of dynamical friction, we refer the reader to \citet{Arcasedda2015,Arcasedda2016,Liu2024}. 

We define a merging radius, $r_{\rm merge}$, so that if a BH wanders inside $r_{\rm merge}$ of the central BH, we assume that the two BHs have merged, updating the mass of the central BH accordingly and removing the merged BH from subsequent tracking. The merging radius is set to 
\begin{equation}
    r_{\rm merge} = \frac{GM_{\rm BH}}{v_{\rm vir}^2}\mbox{\ ,}
\end{equation}
where $v_{\rm vir}$ is the halo virial velocity. We ignore the delay time of BH mergers proceeding under gravitational wave emission, and the gravitational recoil after merger. Thus, we model the optimistic case where BH mergers occur efficiently and do not remove the product from the host galaxy. 

\subsection{Black hole feedback} \label{sec:feedback}

We implement BH feedback through thermal energy injection, resulting from BH accretion. The energy injected to the nearby cold gas at each timestep $\delta t$ is determined as
\begin{equation}
     \delta E_{\rm BH} = \epsilon_w\epsilon_r\dot{M}_{\rm acc}c^2\delta t\mbox{\ ,}
\end{equation}
where $\dot{M}_{\rm acc}$ is the effective BH accretion rate, and $\epsilon_w=0.02$ the radiation-thermal coupling efficiency \citep{Tremmel2017}. This injected energy is converted to an equivalent increase in gas temperature in the vicinity of the BH according to 
\begin{equation}
    \delta T_{\rm feed} = (\gamma-1) \frac{\delta E_{\rm BH}\mu m_{\rm H}}{k_{\rm B} M_{\rm gas}}\mbox{\ ,}
\end{equation}
where $\gamma=5/3$ is the polytropic index of the gas, and $M_{\rm gas}$ the combined mass of the cold and hot gas in the halo. This temperature is used in determining the sound speed for the Bondi-Hoyle accretion rate of Equation~\ref{bondi} (see Section~\ref{sec:accretion} above). 

\begin{figure*}[!htb]
    \centering
    \includegraphics[width=0.9\textwidth]{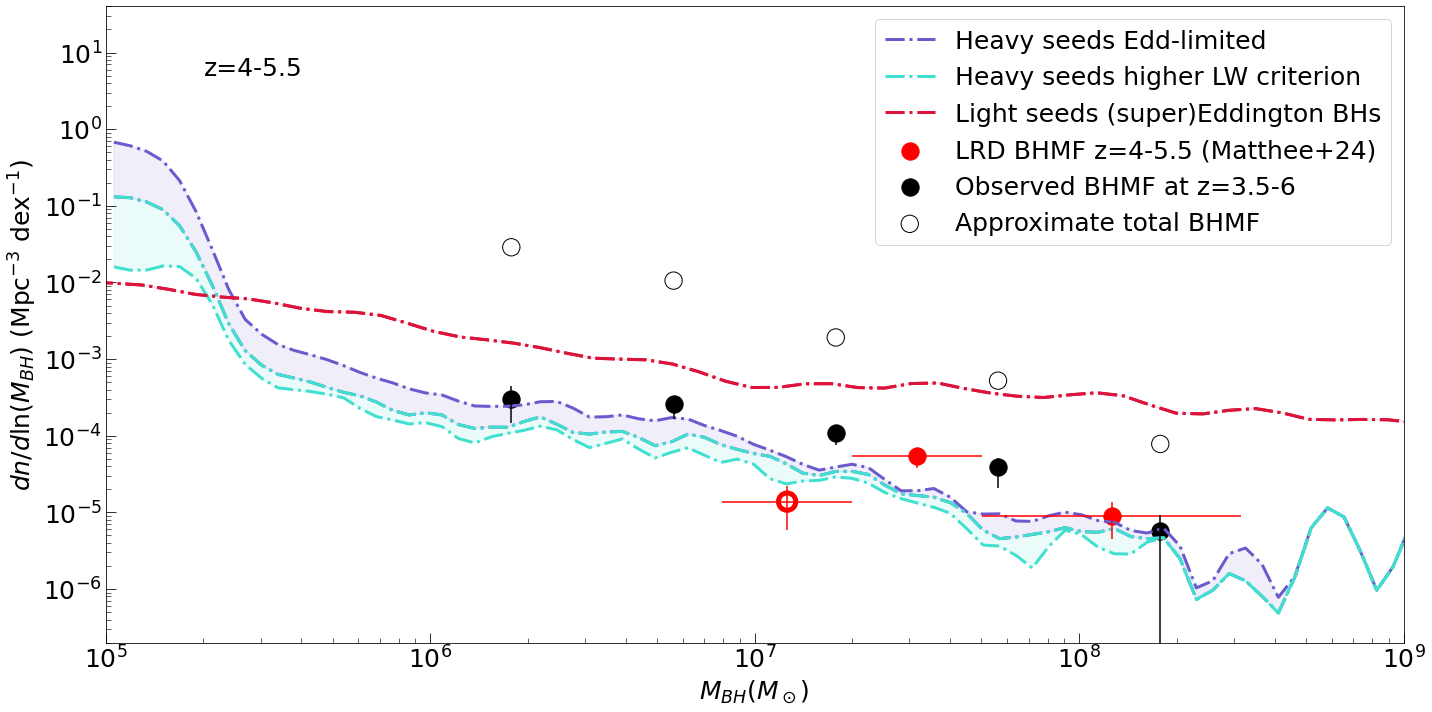}
    \includegraphics[width=0.9\textwidth]{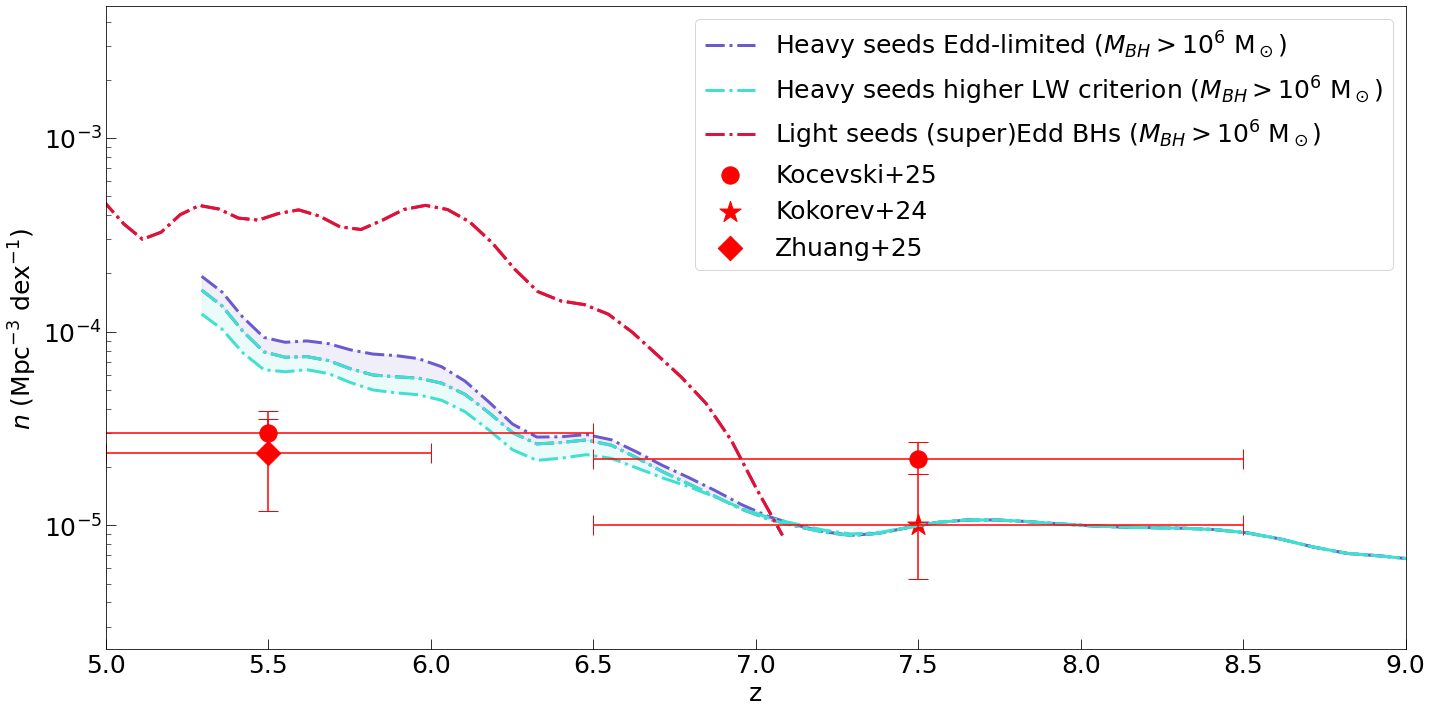}
    \caption{LRD population demographics: mass function at $z=4-5.5$ ({\it top}) and (massive) BH number density  ({\it bottom}). We show models for heavy seeds, heavy seeds with a stricter LW flux criterion for formation, and (super-)Eddington light seeds (i.e., light seeds accreting at or up to 1.5 times the Eddington limit), compared against observations. We include the observed broad line LRD BHMF at $z=4-5.5$ \citep{Matthee2023} and the broad line AGN (BLAGN) BHMF at $z=3.5-6$ \citep{Taylor2024}. We also plot in open circles the approximate total BHMF based on the TRINITY model \citep{Zhang2023}, which includes obscured and dormant SMBHs. The LRD BHMF at the lowest mass bin ($\sim10^7$ M$_\odot$) is shown as an open circle as only three objects were included. For the number density, we include measurements from \citet{Kocevski2025,Kokorev2024_lrd,Zhuang2025}. The heavy seed models largely agree with the LRD observations, but the (super-)Eddington light seed model overproduces the observed BHMF. Furthermore, strengthening the criteria for heavy seed formation does not significantly affect the results. Specifically, applying a higher LW flux threshold does not affect the formation and number density of the most massive DCBH-seeded BHs, which form in the most biased and massive halos (see Fig.~\ref{fig:mhalo}).}
    \label{fig:bhmf}
\end{figure*}

\section{Results}\label{sec:results}

We compare our model outputs with various LRD observations, specifically considering two classes of BH populations: heavy DCBH seeds, and light stellar-remnant seeds that are accreting at Eddington/super-Eddington rates. For the heavy seed models, we consider 4 cases - the Eddington-limited case, super-Eddington limited growth (up to 1.5 times Eddington), forced super-Eddington growth (where all DCBHs accrete at super-Eddington rates when gas is available with $f_{\rm Edd}=1.5$), and the case imposing a higher LW flux threshold for DCBH formation (up to $J_{\rm crit}=3000-10000$). The default heavy DCBH case used for our analysis is the Eddington-limited model, unless stated otherwise. For the light seeds, we consider 2 cases - super-Eddington limited (up to 1.5 times) and forced super-Eddington growth ($f_{\rm Edd}=1.5$). The default model for light seeds is the super-Eddington limited case unless stated otherwise. 

Model details and choices of $f_{\rm Edd}$ and $f_{\rm duty}$ can be found in Table~1 of \citet{Jeon2025}, with the addition here of the higher LW threshold models, which are otherwise identical to the Eddington-limited case. We show that within our model uncertainties, DCBH scenarios agree better with select LRD population statistics, system properties, and spectral energy distributions, compared to the alternative extreme case of super-Eddington light stellar remnant seeds. We note that other recent studies have argued that super-Eddington BHs can explain LRD properties \citep[e.g.,][]{Lambrides2024,Pacucci2024,King2024,Madau2024,Inayoshi2024,Inayoshi2025}, which can be compatible with our work by invoking super-Eddington accretion onto heavy seeds (see Section~\ref{sec:spectra}). We emphasize that LRDs may not originate from a single class of objects or pathways, but through a combination of various origins.

\subsection{DCBH population statistics}\label{sec:pop}

Comparing the population statistics for our BH seed models with the constraints from LRD observations, Figure~\ref{fig:bhmf} summarizes the model predictions for the heavy and light seed populations from the default models compared against the LRD mass function at $z=4-5.5$. We impose the additional criteria to only include light seeds that accrete at Eddington rates or above, labeled ``light seed (super-)Eddington BHs''. As can be seen, the light seed model lies above the observed BHMF, implying that if every (super-)Eddington accreting system became an LRD, their abundance would be overproduced. Conversely, the heavy seed population matches the LRD BHMF well.
Furthermore, we compare the number density of observed LRDs \citep{Kocevski2025,Kokorev2024_lrd,Zhuang2025} against that of massive ($>10^6$ M$_\odot$) BHs in our models. We here only include the more massive objects, reflecting the sensitivity limit of existing LRD observations at $\sim10^6$ M$_\odot$, but this limit may be extended to lower masses in future {\it JWST} surveys that exploit gravitational lensing \citep{Jeon2025}. Similar to the BH mass function, while the heavy-seed population largely matches the LRD number densities, (super-)Eddington light seeds overpredict the observations.

\begin{figure*}[!htb]
    \centering
    \includegraphics[width=0.9\textwidth]{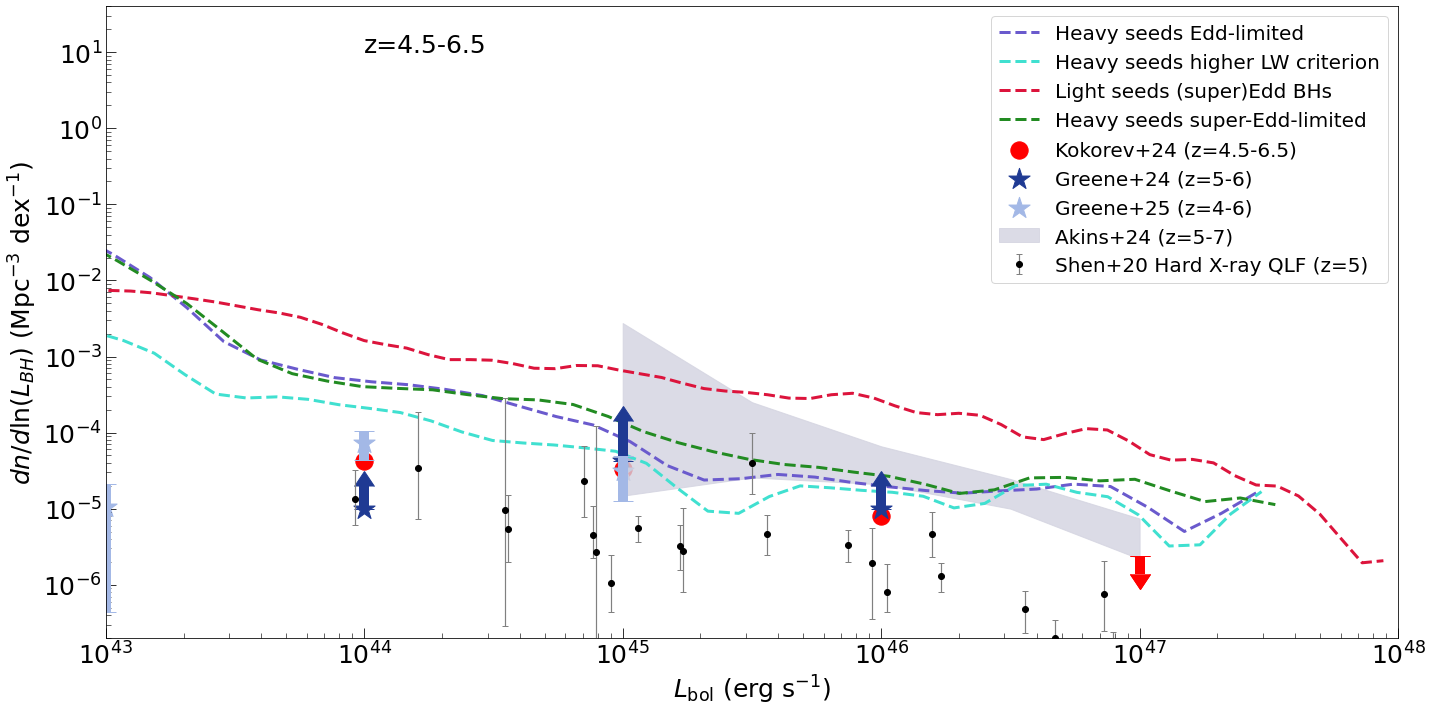}
    \includegraphics[width=0.9\textwidth]{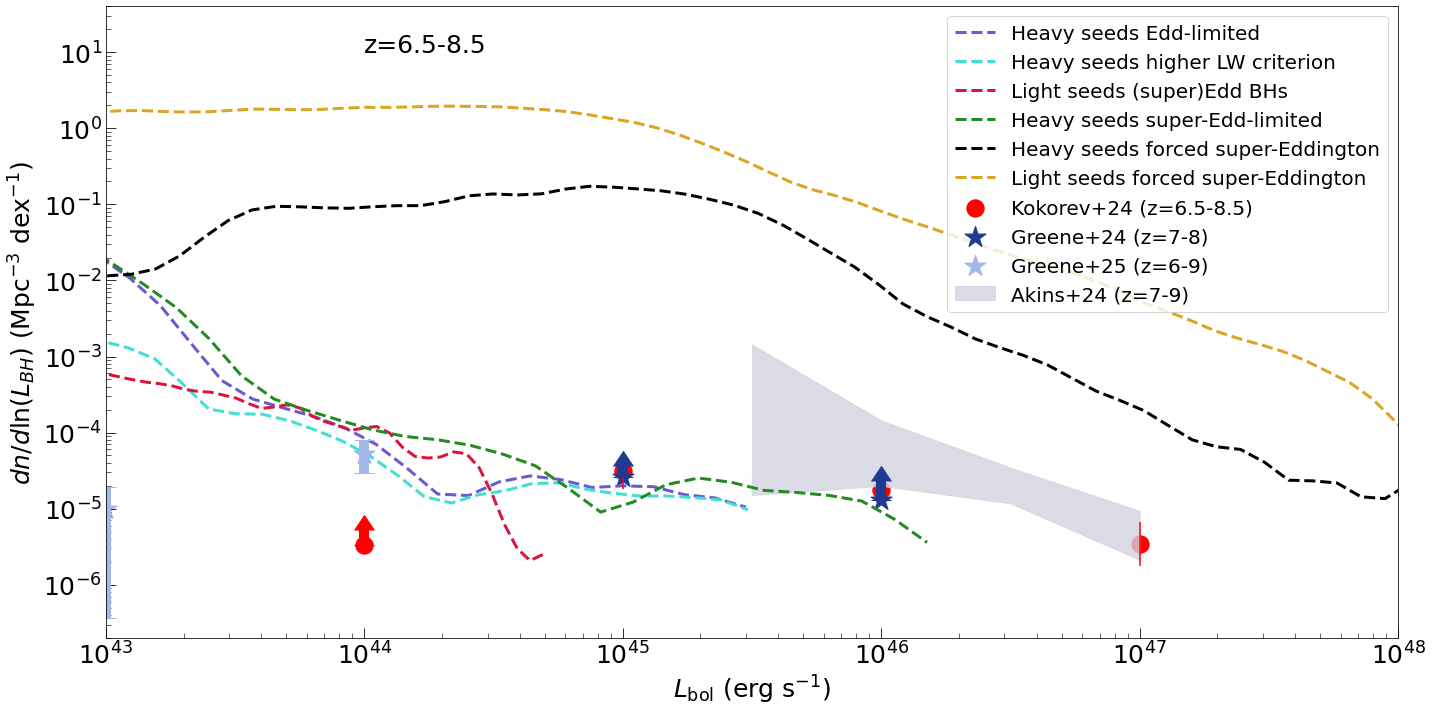}
    \caption{The SMBH bolometric LF models at $z=4.5-6.5$ ({\it top}) and $z=6.5-8.5$ ({\it bottom}), compared with LRD observations. For the lower redshift bin, we consider models for heavy seeds, including the case of heavy seeds with a stricter LW flux criterion for formation, and (super-)Eddington light seeds. We also show the QLF at $z=5$ detected in hard X-rays before \textit{JWST} \citep{Shen2020}. For the higher redshift bin, we additionally show models with forced super-Eddington accretion onto heavy and light seeds. We include various observed LRD LFs \citep{Kokorev2024_lrd,Greene2023,Greene2025,Akins2024}. At $z=4.5-6.5$, the QLF observed prior to \textit{JWST} is lower than the LRD observations, highlighting the unexpected abundance of LRDs. The DCBH heavy seed LF agrees well with the observed LRD LF, while (super-)Eddington light seeds overproduce the LF, similar to Fig.~\ref{fig:bhmf}. However, at $z=6.5-8.5$, while heavy seeds largely agree with the LF shape, both BH seed models cannot reach the highest luminosities. Those luminosities (near $10^{47}$ erg s$^{-1}$) can only be reached by forcing super-Eddington accretion onto heavy or light seeds, which in turn overproduces the LRD LF. This may indicate that a combination of a very small fraction of efficiently accreting heavy seeds are needed to produce the brightest objects, while most of the heavy seed population grows at a slower rate \citep{Jeon2025}. We note that recent work in \citet{Greene2025} corrected their LRD luminosity measurements to lower values by around a factor of 10 in the lower redshift bin and 100 in the higher redshift bin compared to \citet{Greene2023}, so that super-Eddington accretion is not necessary to reach the highest luminosity bin. All models lie above the observed lowest luminosity bin near $10^{44}$ erg s$^{-1}$. This may be due to observational incompleteness for fainter objects, or loss of lower-mass sources due to dynamical effects that are missed in the SAM (see main text).} 
    \label{fig:lmf}
\end{figure*}

However, measuring BH masses for LRDs can be difficult, since spectroscopic observations and relationships between spectral properties and SMBH masses calibrated at lower redshifts need to be used, while their accuracy at higher redshifts is uncertain \citep{Reines2013,Reines2015,Matthee2023,Taylor2024}. In addition, observational incompleteness for fainter LRDs may also affect the observed number density, as reflected in the large measurement uncertainties. Recent observations indicate that LRDs may be more common at lower redshifts than previously thought \citep{Loiacono2025}. This may explain why at lower redshifts ($z\sim5$), the heavy seed population also tends to overproduce the LRD number density. 

A complementary comparison can be made with the LRD bolometric luminosity function (LF), which, assuming a bolometric correction, can be applied to photometrically selected LRDs and thus to a larger sample \citep{Kokorev2024,Akins2024}. In addition, as observations are more sensitive to the brighter objects, the LRD LF observations may be more complete at higher luminosities. Figure~\ref{fig:lmf} compares our model predictions with observed LRD bolometric LFs and the quasar luminosity function (QLF) at $z=5$ detected in the hard X-ray band (2-10 keV), prior to the \textit{JWST} launch \citep{Shen2020}. For our models, we approximately evaluate the bolometric luminosity from SMBH accretion, as follows:
\begin{equation}
    L_{\rm BH} = \epsilon_r\dot M_{\rm acc}c^2\mbox{\ ,}
\end{equation}
with the same definitions of the symbols as above. We specifically carry out comparisons for two different redshift bins, $z=4.5-6.5$ and $z=6.5-8.5$, following \citet{Kokorev2024_lrd}. In the lower redshift bin, the heavy seed LF is largely consistent with various LRD LF observations \citep{Akins2024,Kokorev2024,Greene2023}, whereas the (super-)Eddington light seeds generally lie above the observed values. Both populations overproduce the lowest luminosity bin at $\sim10^{44}$ erg s$^{-1}$ of \citet{Kokorev2024_lrd}. This may be due to observational incompleteness, or the reduction in source abundance, missed in our modeling, when lower mass (and therefore less luminous) SMBHs may wander away from halo centers and thus remain dormant \citep{Jeon2023,Mezcua2020,Reines2020}. Furthermore, the QLF lies below both our models and LRD observations, highlighting the abundance of high-redshift AGN and LRDs, as discovered by \textit{JWST}.

In the higher redshift bin, the trend changes. While heavy DCBH seeds still agree with the observed LRD LF quite well, the subset of (super-)Eddington light seeds from the default model fall below the observations. Evidently, the highest luminosity observations, near $10^{47}$ erg~s$^{-1}$, cannot be reached by either seed model, unless for the extreme case of forced super-Eddington growth. These models assume that every heavy or light seed will grow at 1.5 times the Eddington rate as long as its host halo contains a sufficient supply of cold gas. However, for such extended periods of high accretion, the LRD LF is strongly overproduced. To reconcile these discrepancies, we can consider a scenario where the early Universe contains a general population of heavy seeds, while a very small fraction of them are able to accrete efficiently for extended periods of time, thus being able to produce the highest luminosity objects, without overproducing the number densities \citep[e.g.,][]{Jeon2025}. We note that recent work in \citet{Greene2025} corrected the LRD bolometric LF measurements from \citet{Greene2023} to lower values, by a factor of 10 for $4<z<6$ and 100 for $6<z<9$, so that super-Eddington accretion is not necessary to produce the corrected LF. \citet{Umeda2025} further showed that when applying the dense gas envelope model for LRD spectra (see Section~\ref{sec:spectra}), the effective LRD bolometric luminosity decreases by 1-2 orders of magnitude compared to previous studies that applied dust-reddened spectral modeling. If future work further shows that other LRD bolometric luminosity measurements have been similarly overestimated, extreme super-Eddington accretion may not be required to reproduce the brightest objects.

\begin{figure*}[!htb]
\gridline{
\fig{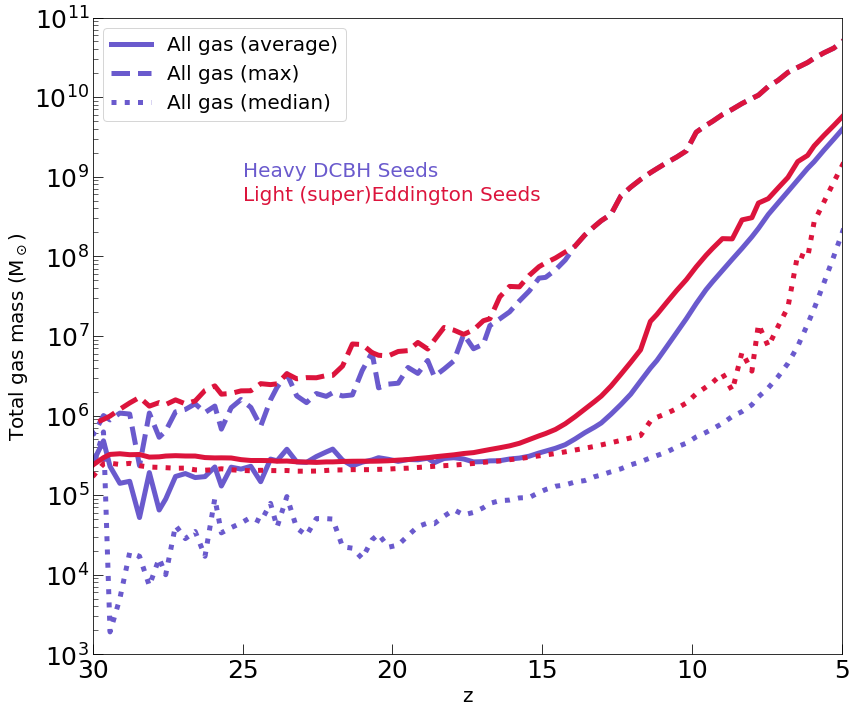}{0.5\textwidth}{}
\fig{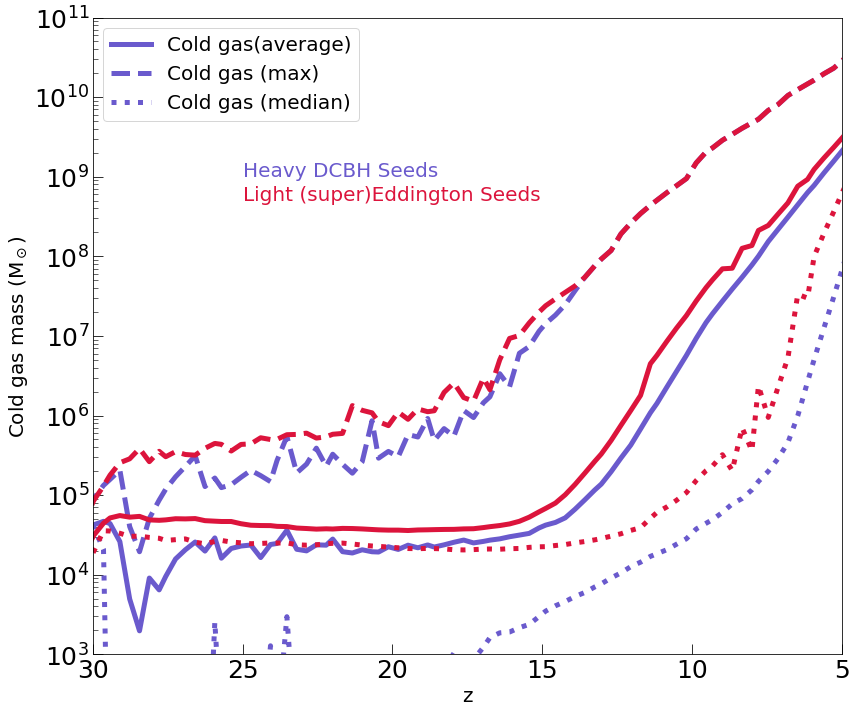}{0.5\textwidth}{}
}
\caption{Global gas supply of LRD hosts. The host halo total ({\it left}) and cold ({\it right}) gas mass, for heavy DCBH and light (super-)Eddington seed models. 
We plot the average, median, and maximum masses for each population of BH seeds. For both the total and cold gas reservoirs, the light (super-)Eddington seed case exhibits slightly higher masses for all categories. Thus, heavy DCBH seeds are more efficient in accreting available cold gas. However, the maximum gas mass for the two kinds of SMBH seeds shows no significant difference, indicating that, rather than the overall gas supply, its distribution near the central SMBH is more important in governing BH growth trajectories, and how extreme objects like MoM-BH* could emerge.}
    \label{fig:gmasses}
\end{figure*}

We note that the heavy seed abundance is not strongly constrained. Although we consider an optimistic heavy seeding model, by changing the criteria for heavy seed formation (e.g., the critical LW flux), their abundance can decrease significantly \citep{Jeon2025}. However, such changes do not affect our overall results. For both Figs.~\ref{fig:bhmf} and \ref{fig:lmf}, we furthermore show cases where we require a higher LW flux to form heavy seeds (also undergoing Eddington-limited growth as in the default model). In comparison to observations, the results are not significantly different from the default model. This is because current SMBH observations capture the more extreme and luminous objects at high redshifts. We show in Appendix \ref{sec:appendix} that currently observed LRDs have estimated host halo masses that match the most massive and biased halos in our models. Therefore, even when tightening the heavy seeding criteria, the most extreme host systems can still form heavy seeds, such that our results at the brightest end are hardly changed. 

Although the above comparison shows that the overall population statistics of DCBHs agree better with LRD observations than light (super-)Eddington BH seed models, this does not rule out the latter as LRD progenitors. As argued above, not every (super-)Eddington stellar remnant BH seed can evolve into an LRD, since that would overproduce the observed population. However, our A-SLOTH accretion model does not change across redshifts, but previous studies have suggested that at lower redshifts, SMBHs may transition into more quiescent accretion \citep{Shankar2009,Ueda2014}. If our models were thus to overestimate SMBH growth in general, the apparent overabundance of light (super-)Eddington BH seeds would not be a problem. The derived abundances, however, should not be largely overestimated, given that our models were calibrated against the high-redshift AGN BHMF \citep{Taylor2024,Jeon2025}, at least in the redshift regime where we make our comparisons. Moreover, a subset of light seeds could still become LRDs, due to select factors not shared by the whole seed population, such as their particular host environments producing the compact and red nature of LRDs. Below, we therefore further compare the properties of the heavy and light seed models with LRD system properties.

\subsection{DCBH system properties}

To further assess the plausibility of DCBH heavy seeds as LRD progenitors, we next consider their host halo's gas content in greater detail. Figure~\ref{fig:gmasses} compares the host halo total and cold gas masses for heavy DCBH and light (super-)Eddington seeds. We show the average, median, and maximum masses for total and cold gas. Somewhat counter-intuitively, we find that, on average, the light seeds have higher total and cold gas masses, especially at early times. This could be due to heavy DCBH seeds being able to accrete the available gas more efficiently, so that at high redshifts, before the halos become more massive, the gas mass is lower in heavy seed systems. We do find that soon after formation, heavy DCBH seeds tend to reside in halos with lower gas mass than light stellar remnant seeds. For the maximum gas masses, no significant difference exists in the two SMBH seed populations. Thus, considering only the overall amount of available gas is not adequate to discern the nature of the BH seed \citep{Jeon2024}, whereas the gas properties near the BH, where dense gas is expected to reside for MoM-BH* like systems, are more crucial.

\begin{figure*}
\gridline{
\fig{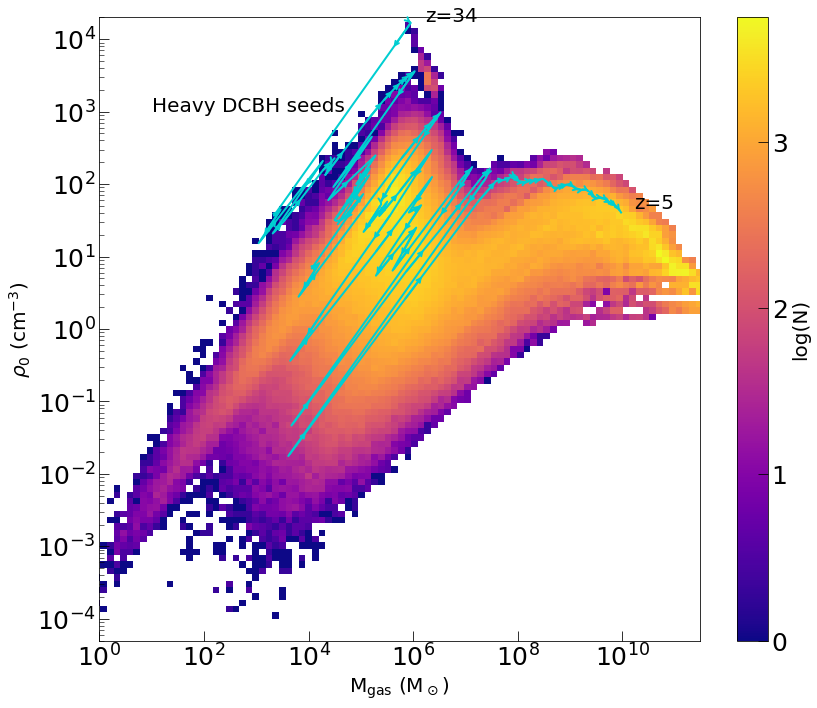}{0.5\textwidth}{}
\fig{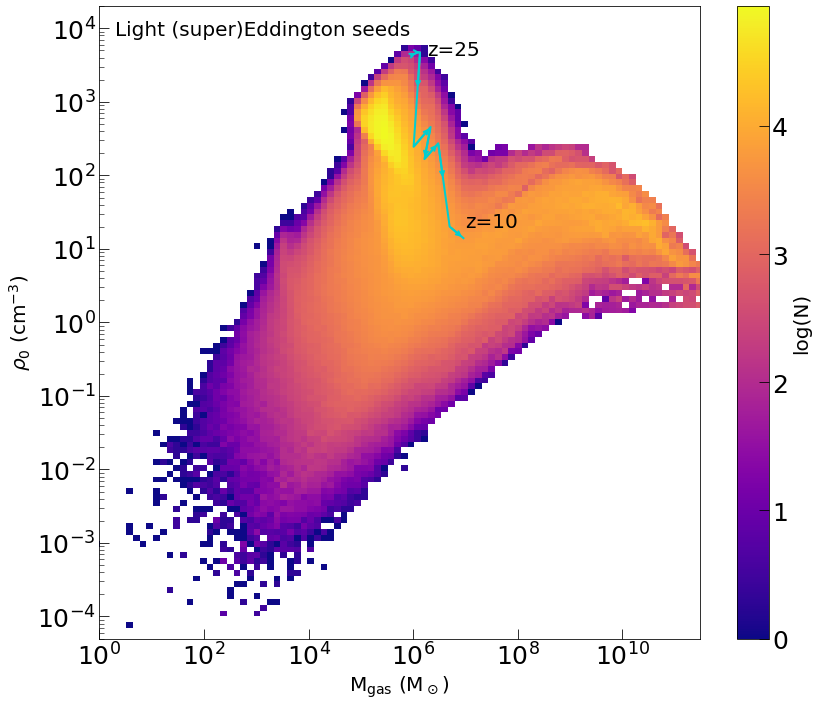}{0.5\textwidth}{}
}
\gridline{
\fig{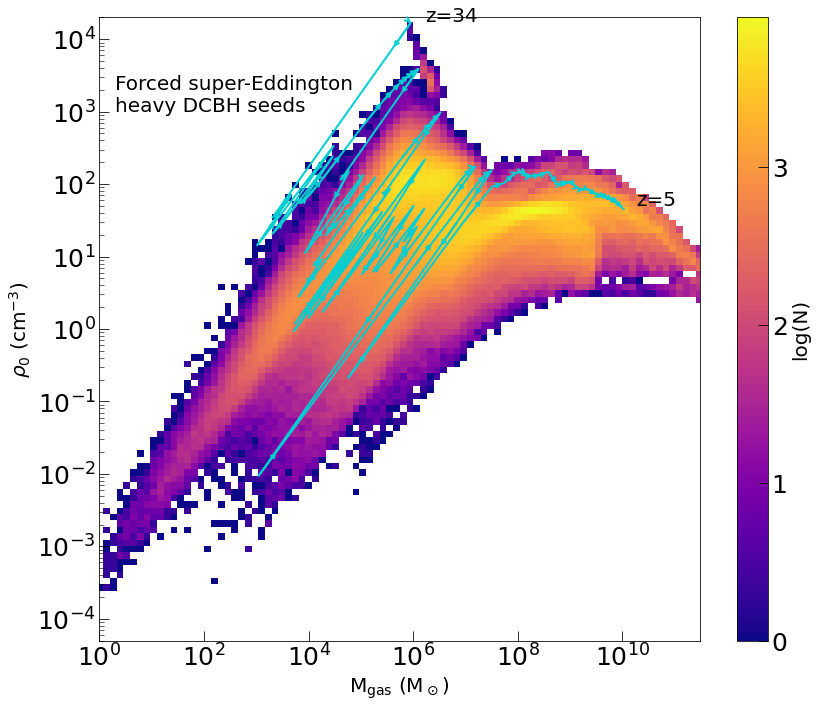}{0.5\textwidth}{}
\fig{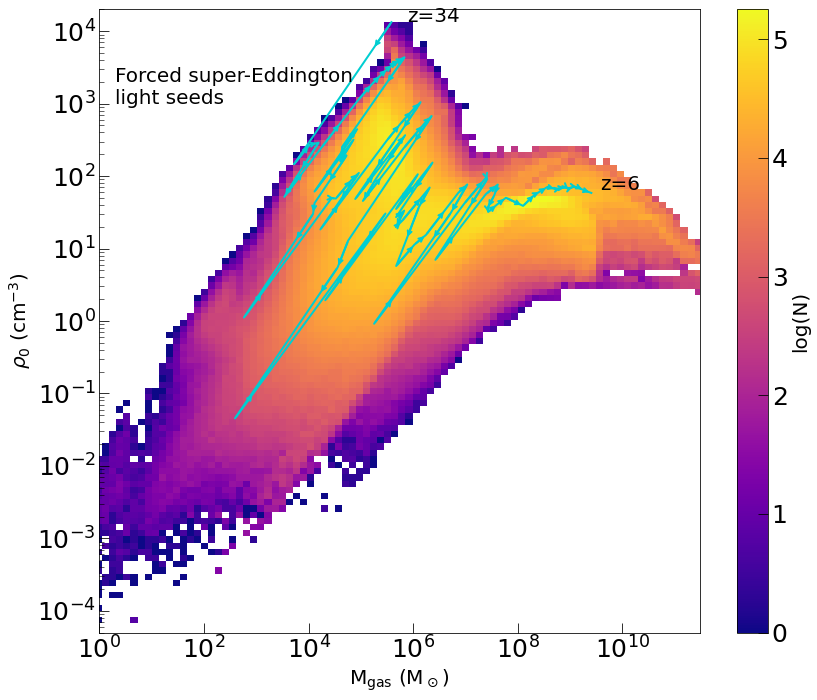}{0.5\textwidth}{}
}
    \caption{Distribution of the central halo gas density $\rho_0$ (assuming an isothermal halo profile) and the total gas mass for heavy seeds, (super-)Eddington accreting light seeds, and forced super-Eddington growth heavy and light seeds. The overall distributions for the two seeds are similar, except that the heavy (DCBH) seed models include systems with a lower central density ($\lesssim10^{-3}$ cm$^{-3}$) at low gas masses ($\lesssim10$ M$_\odot$) and higher central density ($\sim10^{4}$ cm$^{-3}$) at intermediate gas masses ($\sim10^6$ M$_\odot$). We note that the high central density systems ($>10^{3}$ cm$^{-3}$) occur at high redshifts ($z\gtrsim13$). Thus, the presence of heavy seeds can induce high gas densities at early times, but also low densities due to their efficient accretion. We also show in the blue lines the evolutionary tracks of individual systems that reached the highest $\rho_0$ values. The redshifts at the beginning and end of the tracks are indicated. For the more extreme cases of heavy seeds and forced super-Eddington growth, there are rapid phases of gas mass and $\rho_0$ changes throughout the halo's evolution due to gas accretion and outflow. At later times, when the halo is more massive and outflows are not as strong, the track settles down to a phase of increasing gas mass. For light seeds without forced super-Eddington accretion, such rapid changes are not seen, and many objects experience mergers, so that they survive for a shorter period of time. The periods of rapid gas inflow, increasing the central density and gas mass, may be necessary to produce objects like MoM-BH*, which are predicted to be gas rich with low stellar mass at $z\sim8$, long after the primordial conditions of the first BH seed formation.}
    \label{fig:rho0gas}
\end{figure*}

\begin{figure*}
\gridline{
\fig{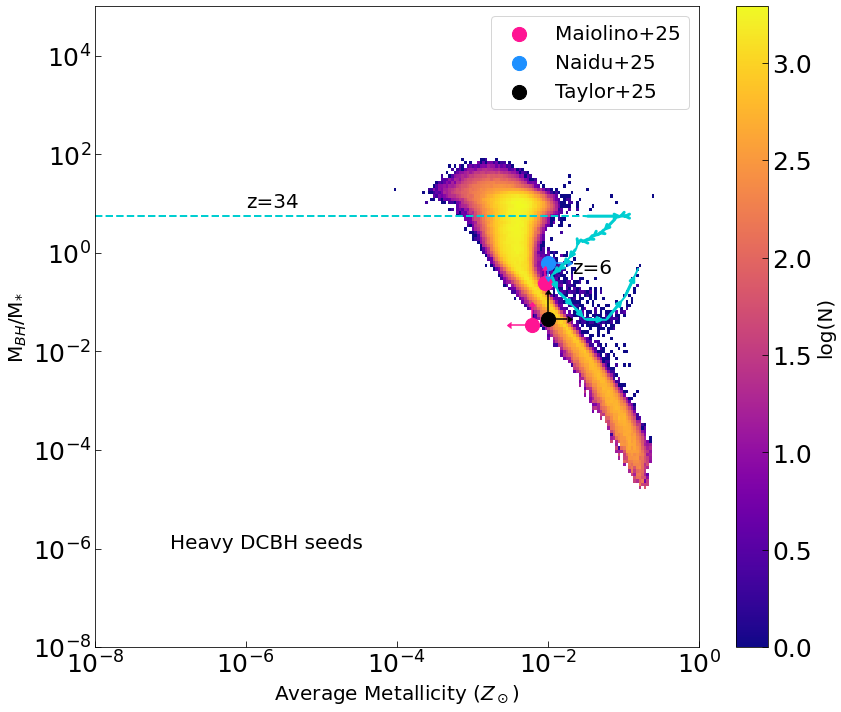}{0.5\textwidth}{}
\fig{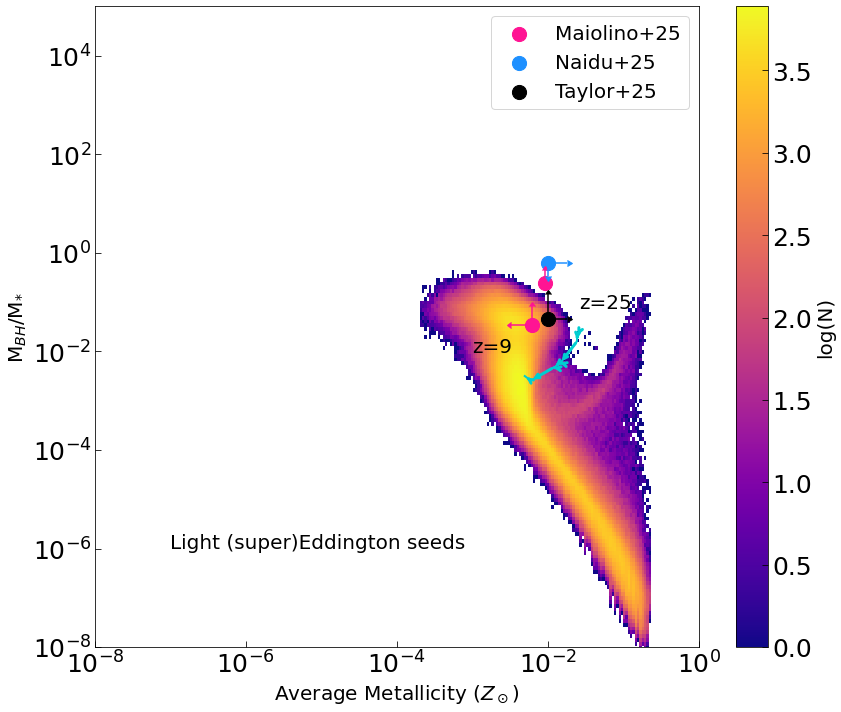}{0.5\textwidth}{}
}
\gridline{
\fig{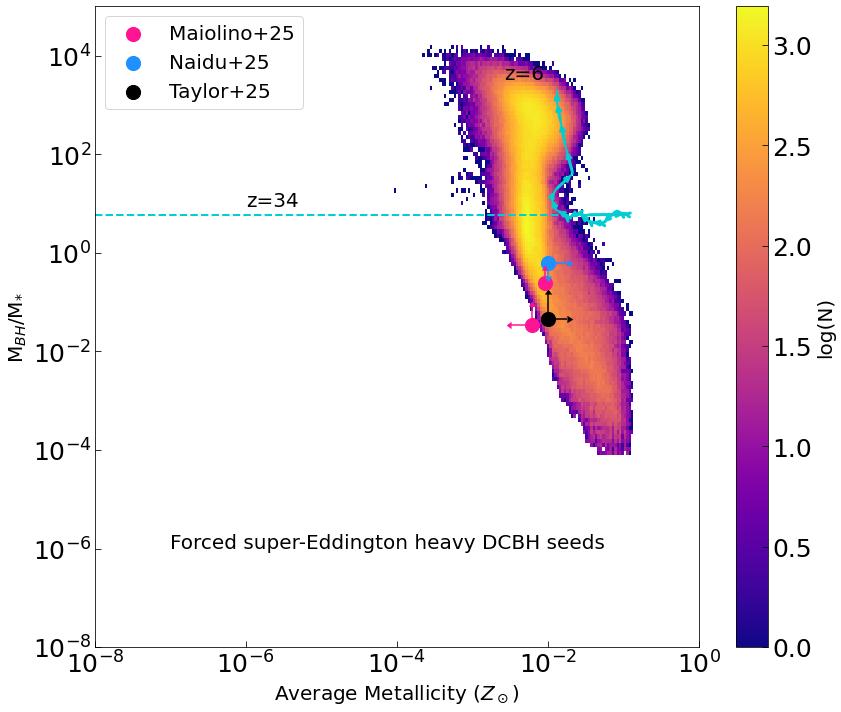}{0.5\textwidth}{}
\fig{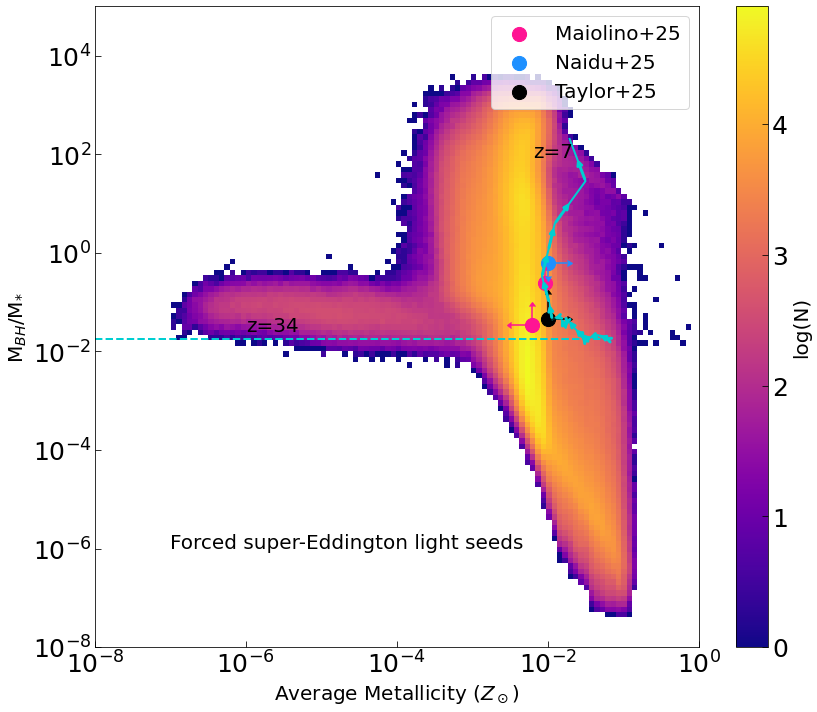}{0.5\textwidth}{}
}
    \caption{Metallicities of SMBH hosts at high-redshift. $M_{\rm BH}/M_*$  vs. the average metallicity for host halos of heavy DCBH seeds, (super-)Eddington light seeds, as well as forced super-Eddington heavy and light seeds for $z=5-10$. For comparison, we show select cases of lensed SMBH observations within this redshift range, Abell2744-QSO1 \citep{Maiolino2025}, MoM-BH* \citep{Naidu2025}, and CAPERS-LRD-z9 \citep{Taylor2025}. While the overall shape of the distribution looks similar for both seed models (other than the light seeds with forced super-Eddington accretion), heavy seeds in general reside in more overmassive systems. Similarly, the observations also prefer heavy seeds to produce overmassive and metal-poor systems (or the extreme case of light seeds with forced super-Eddington growth). Thus, metallicity measurements of high-$z$ AGN could be a key diagnostic of their origins. We also show individual evolutionary tracks for the same objects as in Fig.~\ref{fig:rho0gas} ({\it blue lines}), again indicating their starting and ending redshifts. For the heavy and forced Eddington seeds, they initially form in metal-free systems. We therefore indicate their first timesteps as dashed lines, originating at extremely low (or zero) metallicity values. The chosen tracks represent the extreme systems that reached the highest central density values in Fig.~\ref{fig:rho0gas}.}
    \label{fig:metal}
\end{figure*}

As described in Section~\ref{sec:accretion}, we model the gas distribution in SMBH host halos with a sub-grid approach, as A-SLOTH does not directly provide spatially resolved information within the halos. To assess the differences between the heavy and light seed systems, we specifically approximate the halo gas distribution with Equation~(\ref{isothermal}) and use $\rho_{\rm 0}$ as the gas density at halo center. 

Figure~\ref{fig:rho0gas} shows the distribution of $\rho_0$ vs. halo gas mass for select SMBH seed cases. The heavy seed host halos include both a larger fraction of lower and higher $\rho_0$ systems compared to light seed hosts. The lower $\rho_0$ heavy seed systems correspond to lower gas mass reservoirs, while the higher $\rho_0$ systems show a peak at $M_{\rm gas}\sim10^{6}$ M$_\odot$. For forced super-Eddington growths, there is a larger number of low $\rho_0$ and low gas mass systems, most likely due to the central SMBH efficiently accreting the cold gas. Therefore, while heavy seeds induce a low central gas density for gas-poor systems due to higher accretion efficiency, they can also exhibit high gas densities for sufficiently massive systems. We note that the high-$\rho_0$ cases with $\gtrsim10^3$ cm$^{-3}$ correspond to high redshift $(z\gtrsim13)$ systems, so that at lower redshifts, when the halos are larger, such high densities may not be able to be maintained even in more gas-rich systems.

To see how such a high-density system may evolve, we show individual evolutionary tracks of objects that reached the highest $\rho_0$ value for the four models, respectively. The track for the light seeds is shorter because the less massive stellar remnants experience more merger events and do not survive as long before being incorporated into more massive descendants, as the heavy DCBH seeds. We find that the heavy DCBH seed and forced Eddington growth systems go through phases of rapid gas inflow and depletion. The increase in gas mass occurs through halo accretion and mergers, while the decrease in gas mass is due to multiple factors such as star formation, SMBH accretion, and outflows. At later times, as the halo becomes more massive, the outflows become less strong and the extreme systems settle on an evolutionary track with smaller changes in $\rho_0$ and a consistent increase in the halo gas mass. The rapid changes in the gas mass may be necessary to reproduce gas-rich objects such as MoM-BH* at $z\sim8$, because long after the first BH seeds have formed, the system still holds dense gas with a small stellar mass. In contrast, for light seeds without extreme accretion, the system does not experience such rapid excursions in $\rho_0-M_{\rm gas}$ space, but the gas mass also increases over time, reflecting halo growth. 

To gain further insight into the global properties of the SMBH host systems, for the seeding and growth pathways considered here, we also examine the host halo metallicities. Metal enrichment in A-SLOTH is modeled through supernova outflows, with specific yields linked to the halo stellar population mix (see section~2.5 in \citet{Hartwig2022} for more details). In Figure~\ref{fig:metal}, we show $M_{\rm BH}/M_*$ ratios vs. average metallicities for systems within the  $z=5-10$ range. As expected, heavy seeds form more overmassive systems than light seeds, as DCBHs initially form as massive BHs ($\sim10^5$ M$_\odot$) in star-poor environments. The forced super-Eddington growth cases produce an overabundance of extremely overmassive systems, which likely do not exist in reality \citep{Jeon2025}. The spread in average metallicity is similar between the light and heavy BH seed populations, but heavy seeds show a distinct lack of overmassive systems ($M_{\rm BH}/M_*\gtrsim10^{-2}$) with high metallicity ($Z\gtrsim10^{-2} Z_\odot$), compared to the light seed models. As can be seen, the forced super-Eddington light stellar-remnant systems span a particularly large range in metallicity, as they also sample evolutionary stages with lower metallicity. 

A key lesson here is that metallicity is a promising diagnostic to disentangle the seeding and growth pathways of LRDs and other AGN at high redshifts, in effect providing an additional dimension in the evolutionary parameter space of the first SMBHs. We show in comparison select observations of overmassive broad-line AGN within a similar redshift range, specifically Abell2744-QSO1 \citep{Maiolino2025}, MoM-BH* \citep{Naidu2025}, and CAPERS-LRD-z9 \citep{Taylor2025}. We note that the first example (QSO1) has an unusually low metallicity, challenging all existing models (see \citealt{Maiolino2025} for further discussion), and the metallicity values for the other two examples are taken from photoionization modeling best-fit values. The standard light seeds have difficulty reproducing the observational constraints, not able to produce highly overmassive and metal-poor systems simultaneously, while the heavy seeds and forced super-Eddington accretion can produce such objects. 

We again show the individual evolutionary tracks of the same objects as in Fig.~\ref{fig:rho0gas}. The heavy seed track starts overmassive, then evolves to be less so, but reverses trend and eventually grows to be more overmassive again by $z\sim6$. This is likely due to the accelerated growth experienced by this heavy seed at a rate that is higher than that of star formation at later times. However, only a few extreme cases of heavy seeds follow such a track, like the object represented here, as it formed in the highest central-density region of Fig.~\ref{fig:rho0gas}. The light seed evolves to be less overmassive, but its host system also becomes less metal-enriched on average. This may be due to pristine gas inflow to the halo. For the forced super-Eddington case, the system grows to be extremely overmassive, as expected for extended periods of efficient SMBH accretion, but such extreme growth may only be realized for rare cases.

From the discussion above, we further note that when we force the BH seeds to grow at a super-Eddington rate, the light seeds can also produce the extreme overmassive and metal-poor systems, making the distinction with heavy seed origins difficult. This arises because the cold gas, which enables BH accretion, is also the fuel for star formation. To sustain super-Eddington accretion, the cold gas in the halo must be provided to the BH, which in turn leaves little gas for stars to form and enrich the halo, resulting in a system with a massive SMBH, but low stellar mass. However, the light seed system at first is not extremely overmassive with $M_{\rm BH}/M_* \sim 0.01$, as a previous stellar population must exist unlike the heavy DCBH systems which can be extremely overmassive ($M_{\rm BH}/M_* \gtrsim 1$) soon after formation. Therefore, searching for fainter objects (at earlier evolutionary stages) is important to distinguish different SMBH formation channels \citep{Jeon2025}.

In the context of LRDs, the MoM-BH* object has a low inferred metallicity ([Fe/H]$\sim -2$), based on photoionization modeling \citep{Naidu2025}. If MoM-BH* is an AGN with no significant population of stars, this is consistent with the heavy DCBH scenario. Super-Eddington light seed models, while being able to produce the metal-poor and overmassive system, would still have difficulty in producing a near-starless system where the total emission is dominated by the AGN (see Section~\ref{sec:spectra}), given that Pop~III stars must have formed prior to the emergence of a stellar remnant. Thus, systems like MoM-BH* may more naturally arise from a heavy seed origin, while presenting remaining challenges to all models, including the DCBH one.

\subsection{Model spectra}\label{sec:spectra}

\begin{figure*}[!htb]
    \centering
    \gridline{
\fig{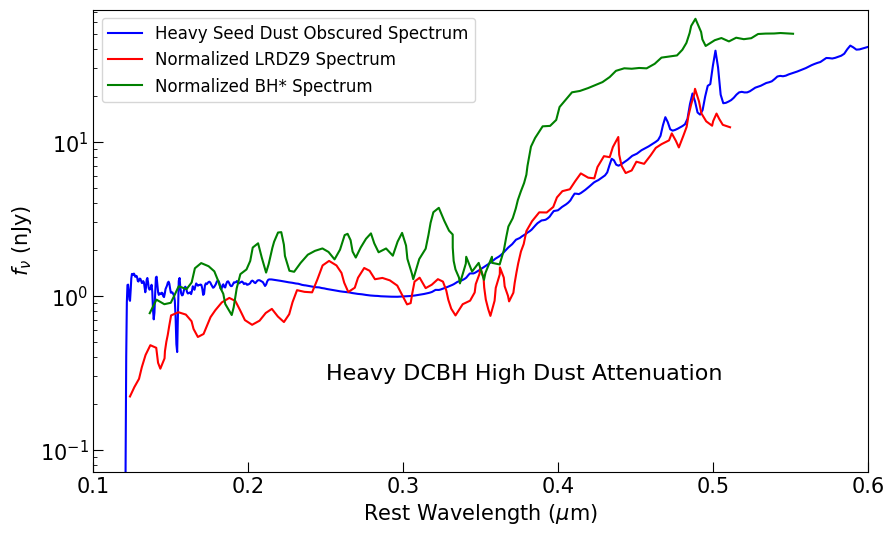}{0.5\textwidth}{}
\fig{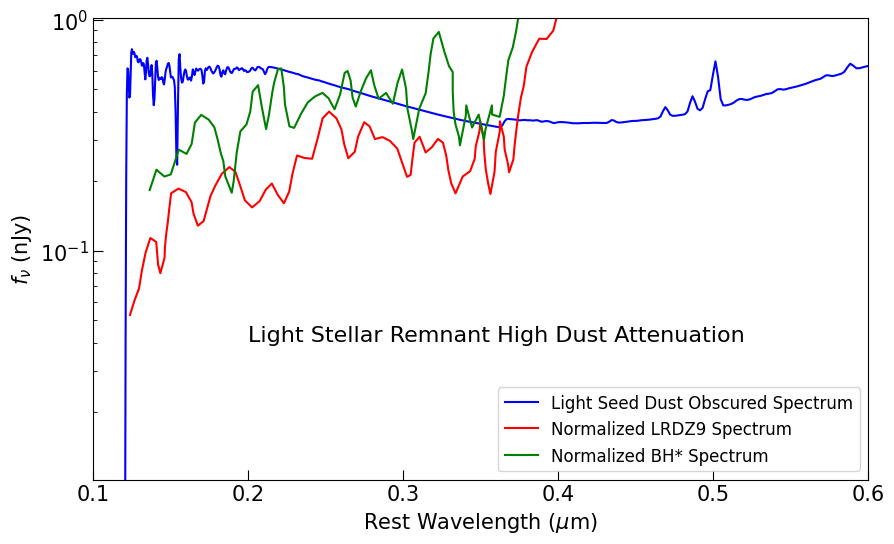}{0.5\textwidth}{}
}
    \gridline{
\fig{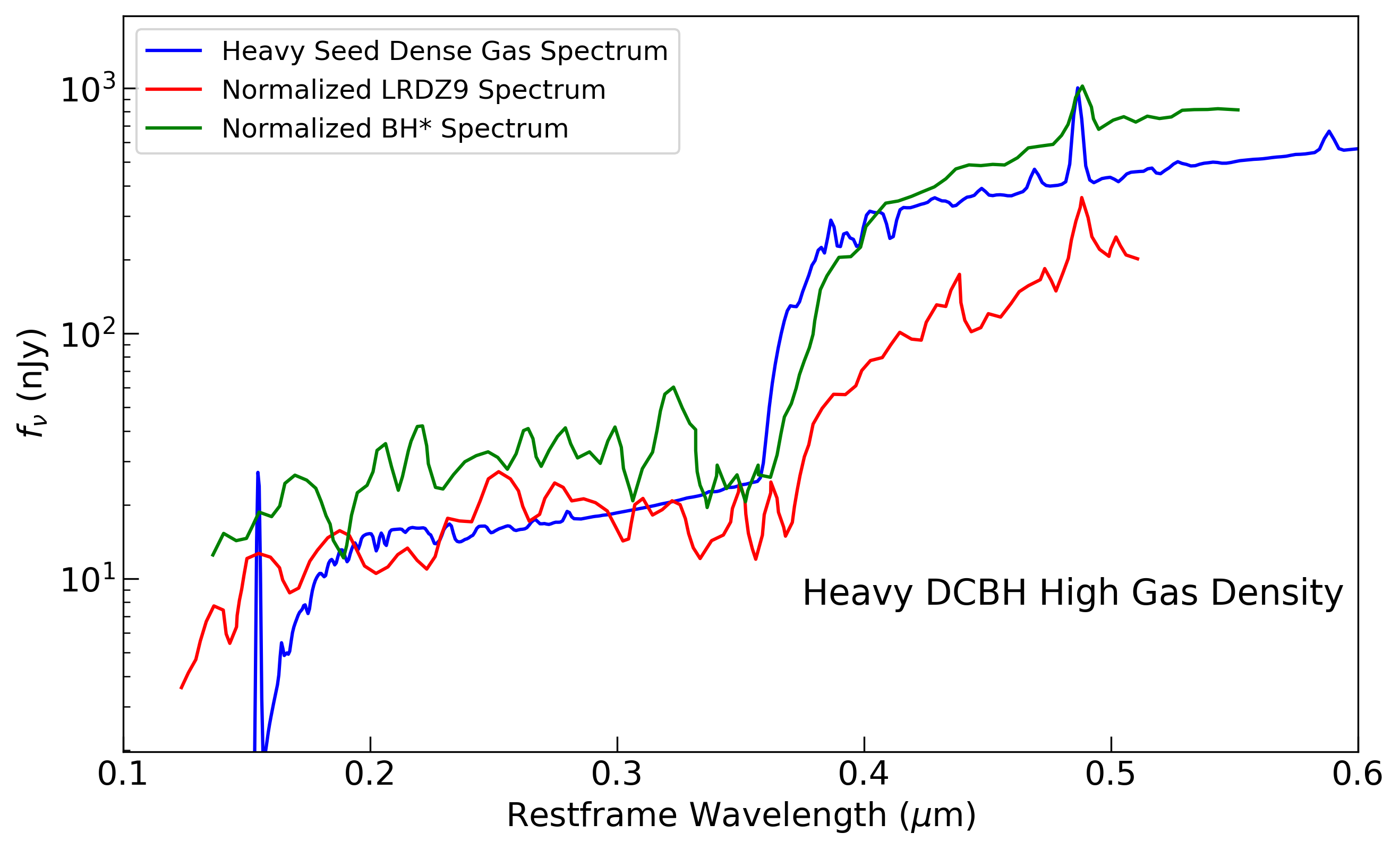}{0.5\textwidth}{}
\fig{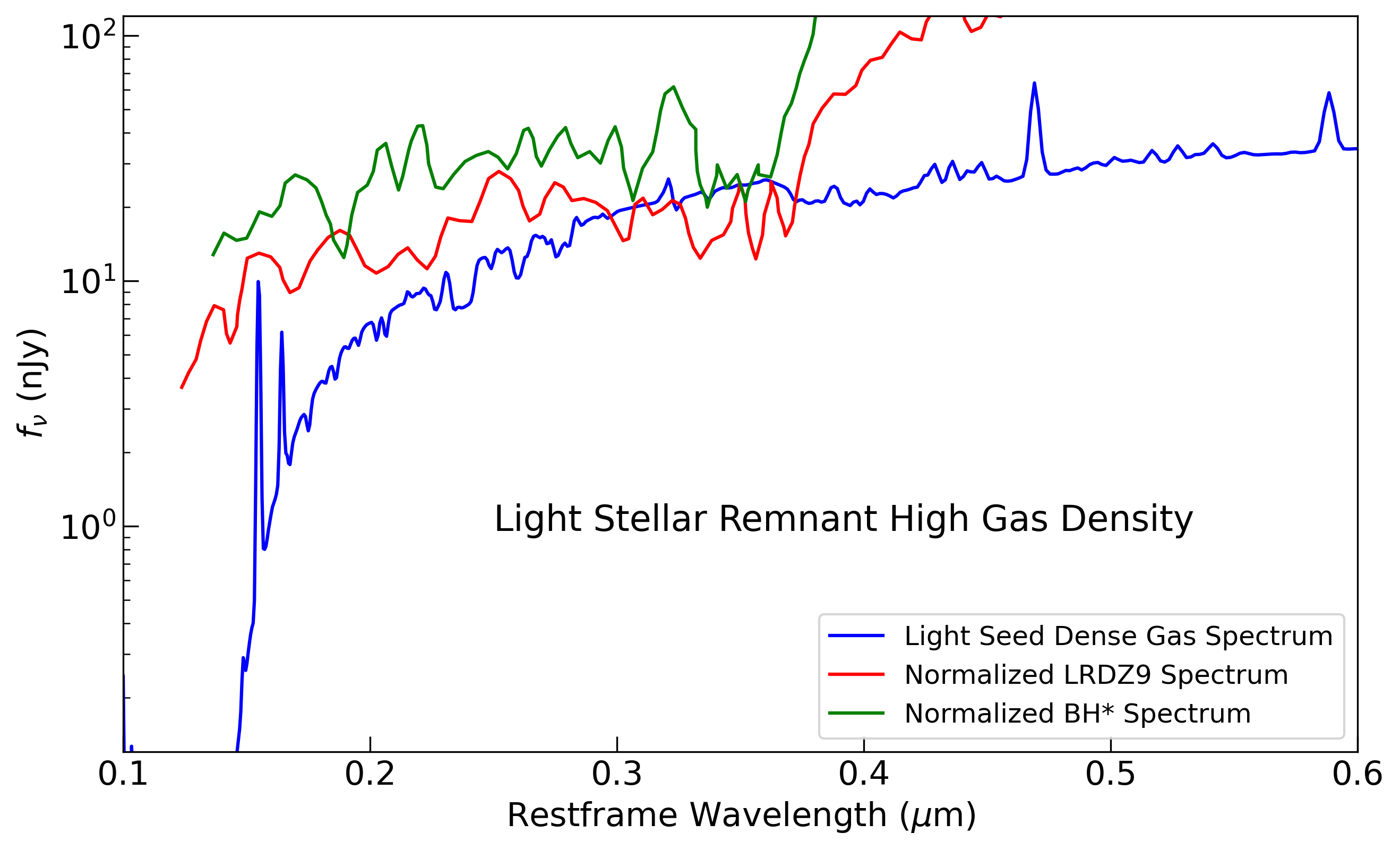}{0.5\textwidth}{}
}
\caption{Synthetic spectra produced with Synthesizer \citep{Lovell2025,Roper2025} under high dust attenuation ({\it top}), and with \textsc{Cloudy} \citep{Chatzikos2023} under dense gas conditions ({\it bottom}) for our A-SLOTH SMBH seed models. We produce spectra at $z=10.1$ for the extreme cases of forced super-Eddington growth onto the BH seeds for extended periods, to compare against select extreme objects like CAPERS-LRD-z9 \citep{Taylor2025} and MoM-BH* \citep{Naidu2025}. We consider both the heavy and light seed cases under the same assumed environmental conditions (see Section~\ref{sec:spectra}), and compare them against the normalized spectra (at 0.35 \micron) of CAPERS-LRD-z9 and MoM-BH*. The heavy seed, with its higher mass ($\sim3\times10^8$ M$_\odot$) and luminosity, is able to produce the distinct red shape of the BH*-like objects and the strength of the Balmer break under both dust attenuation and dense gas scenarios. In contrast, the light seed ($\sim3.5\times10^6$ M$_\odot$), even if it is accreting at super-Eddington rates, cannot produce such extreme red spectra with strong Balmer breaks. Therefore, for LRDs to exist at $z\sim9-10$ as observed, and if BH*-like objects are progenitors of LRDs \citep{Naidu2025}, the super-Eddington heavy seed scenario offers a more compelling origin, as it is able to reproduce the red spectra more easily than light seed models, which predict AGN that are not as massive and luminous at early times.} 
    \label{fig:spectra}
\end{figure*}

As an additional signature, we derive select model spectra, based on the A-SLOTH outputs, to compare against high-redshift LRD spectroscopic observations with {\it JWST}. We first use Synthesizer \citep{Lovell2025,Roper2025}, designed to produce synthetic spectra from simulations and SAMs. Within Synthesizer, we use \textsc{Cloudy} version C23.01 \citep{Chatzikos2023} for photoionization modeling. From A-SLOTH, we chose the extreme cases of forced super-Eddington growths onto heavy and light seeds to compare against extreme objects like CAPERS-LRD-z9 \citep{Taylor2025} and MoM-BH* \citep{Naidu2025}, inferred to be AGN embedded in dense gas with limited stellar components. We specifically consider the most massive SMBH at $z\simeq10$ to match the high redshift of CAPERS-LRD-z9 and considering that current high-redshift observations are likely biased towards the most massive/brightest BHs \citep{Lauer2007,Li2025}, using the modeled SMBH properties and the host galaxy stellar mass as inputs. The heavy seed has the mass $M_{\rm BH}\sim3\times10^8$ M$_\odot$ and the light seed $M_{\rm BH}\sim3.5\times10^6$ M$_\odot$. For the AGN spectra, we utilize the QSOSED spectral model \citep{Kubota2018}, and for the stellar spectra the Binary Population and Spectral Synthesis (BPASS, version 2.2.1) package \citep{Stanway2018,Eldridge2017}. We consider both Pop~III and II stellar populations within the SMBH host system. For Pop II stars, we assign their metallicity to the halo average metallicity, whereas for Pop~III, we assume a near-zero metallicity value. 

We infer the conditions of the interstellar medium (ISM), including any dust attenuation, near the SMBH and within its host galaxy with a sub-grid approach (see Section~\ref{sec:accretion}), as they are not directly provided with SAM outputs. We initially consider the high dust attenuation scenario proposed to explain the red spectrum (in the rest-frame optical) of the LRDs \citep{Durodola2024,Huang2024,Kokorev2024}. For ISM dust attenuation, we assume a large V-band value of $A_V =5$, with a power-law extrapolation according to $A_{\lambda}\propto \lambda^{-1.3}$. The metallicity for the selected systems is low $(Z\sim10^{-1.5}~Z_\odot)$, so to achieve such high dust attenuation, we assume that their galaxies are extremely compact to have high dust concentration following the LRD observations \citep{Baggen2024}. Furthermore, we assume a blackbody dust emission with a temperature of 100~K. We further assume that the stars experience attenuation from their birth clouds \citep{Charlot2000}. We perform spectral modeling for the light and heavy seed scenarios under the same assumptions regarding gas, ISM, and dust properties. 

The resulting spectra are shown in Figure~\ref{fig:spectra}. First, exploring the case of significant dust attenuation ({\it top panels}), we have chosen the free environmental parameters (most importantly, the dust attenuation and emission) so that the heavy seed case is able to reproduce the observed spectral shape of high-redshift LRDs. The distinctly red spectral shape of CAPERS-LRD-z9 and MoM-BH*, including its strong Balmer break, can be reproduced by the dust-attenuated heavy seed model, when normalized to match the synthetic spectrum. Here, the red spectral portion is produced by the highly-accreting SMBH, whereas the bluer part of the spectrum is generated by the stellar population. In contrast, the light seed is not luminous enough, compared to its host's stellar population, so that the spectrum is not as extremely red and thus cannot reproduce the observations. 

Next, we test the second scenario proposed to explain the red LRD spectra, involving dense gas surrounding the AGN \citep{Naidu2025,Inayoshi2025_bl}. For this case, we directly use \textsc{Cloudy} to model the spectrum. In addition, we use Yggdrasil \citep{Zackrisson2015} to model the Pop~III component, as \textsc{Cloudy} does not directly vary the metallicity of the stellar population. The AGN and Pop~II input spectra remain the same as before\footnote{As Pop~III is subdominant to the more massive Pop~II stellar component, this change in the input spectral library has no significant impact on the resulting spectrum.}. In the dense gas scenario, following \citet{Inayoshi2025_bl}, we assume that high-density gas of $10^{8}$ cm$^{-3}$ surrounds the central AGN, with contributions from nebular emission with a covering fraction of 0.9. Such a gas density is much larger than the derived $\rho_0$ values presented in Fig.~\ref{fig:rho0gas}, where we assumed an isothermal halo gas profile. Therefore, in this case we implicitly assume that the small-scale gas density near the central SMBH is much higher than the larger-scale, core gas density of the halo. We also include a turbulent gas velocity of 300 km s$^{-1}$, and low dust attenuation following the Small Magellanic Cloud (SMC) reddening law \citep{Gordon2003,Taylor2025} with $A_V=0.5$. As in the high-dust attenuation scenario, we apply the same gas conditions for the heavy and light SMBH seeds. The bottom panels of Fig.~\ref{fig:spectra} show the resulting \textsc{Cloudy} spectra. Again, the heavy seed case is able to reproduce the red spectrum and strong Balmer break of LRD observations. In contrast, the light seed models, involving an AGN component that is too faint, are unable to reproduce these characteristic LRD spectral features. 

Consequently, for the observed LRDs at $z\sim9-10$, and if MoM-BH*-like objects are progenitors of LRDs \citep{Naidu2025}, the heavy seed scenario provides a more robust explanation compared to the light stellar remnant seeds, even if they are accreting at super-Eddington rates. We acknowledge that there are multiple free parameters involved in the spectral modeling, especially related to environmental conditions, that can be varied to produce different spectral shapes. Specifically, we require high dust attenuation or high gas density to produce such a red spectrum and strong Balmer breaks. Under sustained super-Eddington accretion over extended periods, as assumed in our forced super-Eddington cases, it is not known how likely these environmental conditions are. Furthermore, the stellar component could be made fainter, if the galaxy is in a quenched phase, which may be possible even at high redshifts \citep[e.g.,][]{Weibel2025}, or if dust obscures the galaxy while not affecting the central AGN as strongly \citep{Fujimoto2022}. Then, the less bright light stellar remnant seeds could still reproduce the red spectra. Therefore, our results indicate that the observations of select high-$z$ LRDs could be explained under the extreme conditions of super-Eddington accretion onto heavy seeds, and of the presence of the dense gas environment that supports such extreme accretion\footnote{If DCBHs initially form in starless/star-poor systems, super-Eddington accretion may not be necessary to produce the red spectrum, as the DCBH will not have to outshine the existing stellar population, different from the case of stellar-remnant BHs.}.


\section{Feasibility of DCBH Pathway}\label{sec:discussion}

We have argued above that the DCBH population can better explain aspects of the observed LRDs, and we will here consider the broader context for this scenario. Assuming that they are indeed powered by AGN, LRDs are a subset of the total observed AGN population \citep{Taylor2024,Maiolino2023}. Recent work further showed a bimodality of BH-to-stellar-mass ratios in the observed broad-line AGN, where the AGN characterized as LRDs were overmassive while non-LRD AGN were not \citep{Jones2025}. Thus, it would not be surprising if LRDs similarly originated in a subset of all SMBH seeds, such as the DCBH class which can easily produce overmassive systems. In support of this idea, we recall that Figs.~\ref{fig:bhmf} and \ref{fig:lmf} show that the heavy seed population agrees better with select LRD population statistics, including number density, BHMF, and LF, compared to the more common and numerous highly-accreting light seed SMBHs. We emphasize again that if all (super-)Eddington accreting light seeds were to form LRDs, the observations would be overproduced, while the less common heavy seed systems agree surprisingly well with the observed LRD demographics. 

\begin{figure*}
\gridline{
\fig{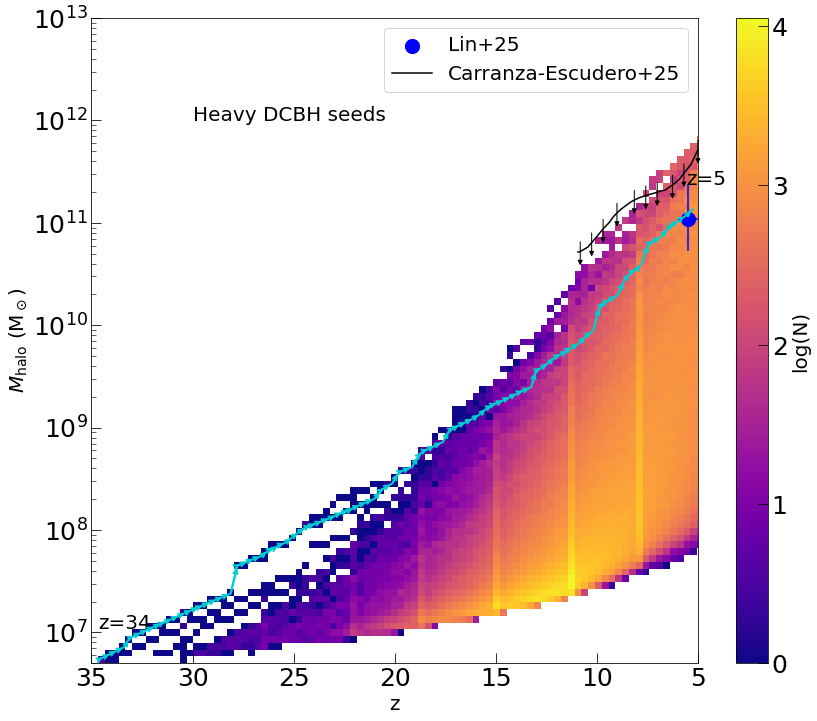}{0.5\textwidth}{}
\fig{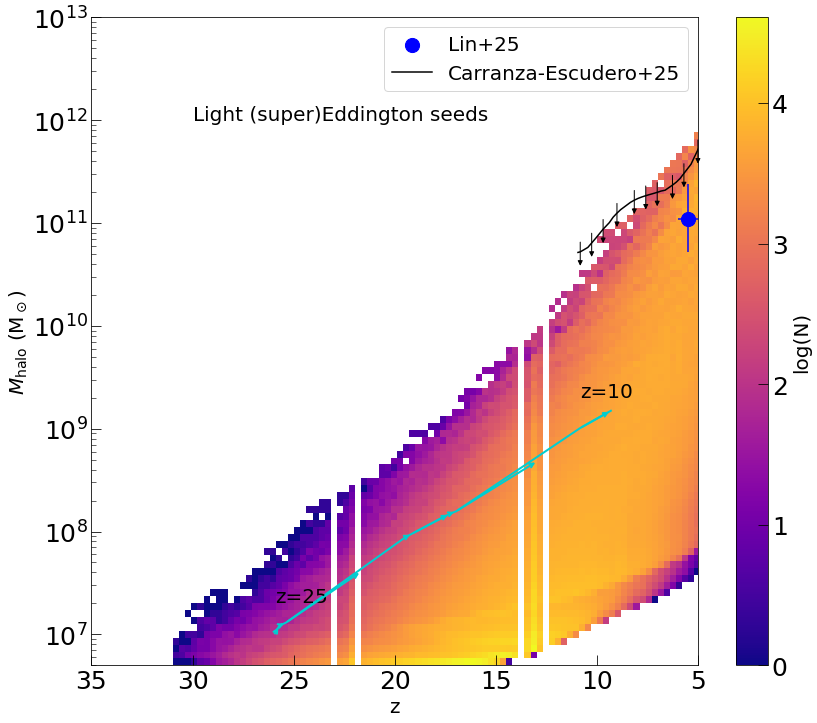}{0.5\textwidth}{}
}
\gridline{
\fig{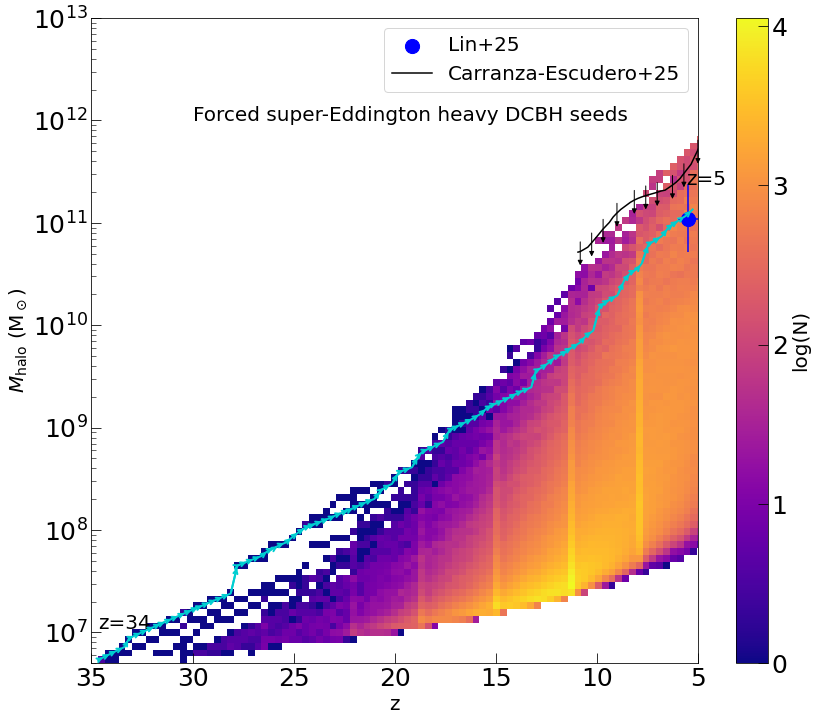}{0.5\textwidth}{}
\fig{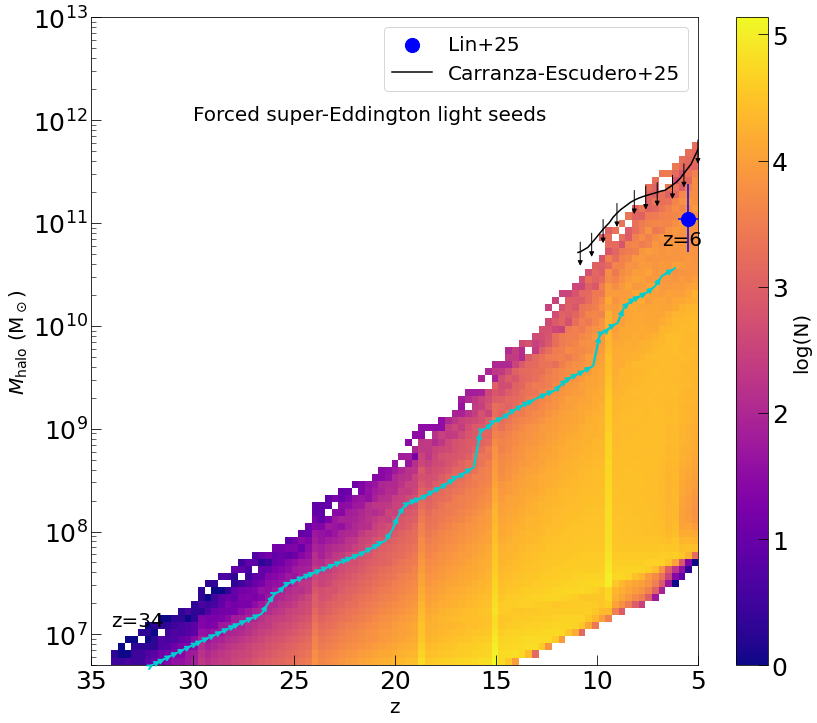}{0.5\textwidth}{}
}
    \caption{Host halo mass against redshift, $M_{\rm halo}$ vs. $z$, for heavy DCBH seeds, (super-)Eddington light seeds, as well as forced super-Eddington heavy and light seed models. In comparison, we show the estimated halo masses of LRDs from abundance matching and clustering analysis \citep{Lin2025,Carranza2025}. We also show the select individual tracks that reached the highest central density values in Fig.~\ref{fig:rho0gas}. In general, the heavy DCBH population resides in higher mass halos ($\gtrsim10^7$ M$_\odot$) compared to the light stellar seed hosts, even at higher redshifts $(z\gtrsim15)$. All models can match the current observational constraints on LRD host halo masses, which align with the more massive and biased population of halos at $\sim10^{11}$ M$_\odot$. We thus conclude that current observations identify the most biased hosts for LRDs.}
    \label{fig:mhalo}
\end{figure*}

When considering the overall gas supply, all models exhibit a very similar behavior, whether hosting heavy or light seeds (as summarized in Figs.~\ref{fig:gmasses} and \ref{fig:rho0gas}), with the slight exceptions discussed above. Similarly, for the halo masses, all models span a wide range (Fig.~\ref{fig:mhalo}). However, there is a clear difference in the tracks for individual systems that reached highest gas densities, comparing heavy and light seeds, if the latter do not experience forced super-Eddington growth (see Fig.~\ref{fig:rho0gas}). We note that the large excursions in gas supply found for select model galaxies early in their evolution could account for objects like MoM-BH* and CAPERS-LRD-z9, with their inferred sub-dominant or absent stellar components. Given that they are observed at a time much later than what is expected for first star formation $(z\sim30)$, such rapid resupply of gas during a subsequent stage may explain their properties. In such a setting, star formation may be hampered by the highly accreting AGN, even as the halo accretes additional gas from the cosmic environment. Furthermore, no previous history of significant star formation would be necessary to produce an AGN, unlike in the case of light stellar remnant seeds experiencing less aggressive growth (see top-left panel in Fig.~\ref{fig:rho0gas}). 

We note that when super-Eddington accretion onto the BH seeds is forced throughout their evolution, as opposed to allowing the accretion to react to changing gas conditions in the host centers, effectively implying a 100\% duty cycle, the halo gas properties discussed above could also be reproduced by the light seed scenario. This highlights the need to further constrain the fraction of BHs that are accreting at super-Eddington rates, together with the corresponding duty cycles for heavy and light seeds, both from the observational and theory/simulation side. For instance, it is shown by \citet{Kiyuna2025} with analytic arguments supported by toy-model simulations that rapid growth of light seeds stalls at $\sim 10^4\ \rm M_\odot$ due to the forward acceleration effect caused by the ionized bubble around the BH. 


Another promising observational indicator of the physical nature of LRD progenitors, regarding the SMBH seeding and growth pathway, is their metallicity (see Fig.~\ref{fig:metal}). Heavy DCBHs are born overmassive with a massive SMBH and little to no stellar mass. In contrast, light stellar remnant BH systems already host a stellar population. Thus, these models experience markedly different metal enrichment histories, where heavy seed systems exhibit extremely overmassive and metal-poor configurations not seen in light seed models. 
The recently observed Abell2744-QSO1 lensed AGN at $z=7.04$ \citep{Maiolino2025} is an intriguing test case, as QSO1 is highly overmassive while also showing near-pristine metallicity. 

This combination of properties challenges both heavy and light seed models\footnote{When comparing specific observations with our model predictions in Fig.~\ref{fig:metal}, one should keep in mind that the figure presents cumulative properties over a significant redshift range, and is not restricted to the corresponding redshift of a given observation. Any overlap between symbols and model contours in the figure may thus be spurious, as they show the properties at different times.}, motivating the consideration of an alternative PBH origin, at least for rare cases such as QSO1 (see the above reference for a detailed discussion).  
We note that in our modeling, a DCBH seed forms through the direct collapse of the progenitor SMS, without releasing any metals into the surrounding \citep[e.g.,][]{Shibata2002_DCBH,Reisswig2013_DCBH}. The heavy-seed metallicities may thus be higher than what is predicted here, if the SMS progenitor were to contribute to local SN enrichment \citep[e.g.,][]{Nandal2024_SMS}. We also assume that there is no explosion during the collapse, while simulations have shown that rotating SMSs can undergo powerful explosions by shock heating driven from a torus around the DCBH \citep{Fujibayashi2025}, or via general-relativistic instabilities \citep{Chen_SMS2014}. Such explosions may temporarily quench star formation and BH growth, prolonging the starless/star-poor phase.

In interpreting our results, a number of caveats should be borne in mind. Our conclusions are based on the A-SLOTH SAM, so that our predictions depend on the accuracy of the SAM prescriptions and their (empirical) calibration. We use an optimistic heavy seeding scenario \citep{Jeon2025} that may inflate their number. However, for the existing LRD observations, only the most massive end of the DCBH distribution is observed, where resulting number densities do not strongly depend on the detailed DCBH seeding conditions (see Section~\ref{sec:pop}). Moreover, we use EPS trees for our models due to computational efficiency, while using halo merger trees from full cosmological N-body simulations may produce more accurate halo statistics and merger histories. In our models, however, SMBHs mainly grow through accretion and not mergers, such that approximating halo merger trees with the EPS formalism is justified for our heuristic exploration of parameter space. 

For our accretion modeling, we use an optimistic duty cycle of $f_{\rm duty}$=0.5 or 0.8. While the observed values for AGN at high-$z$ are significantly lower \citep[e.g.,][]{Arita2025,Pizzati2025,Eilers2024}, the observations define the duty cycle as the fraction of active SMBHs at any given time. The duty cycle in our model instead represents the \textit{maximal} active fraction over the SMBH lifetime (see Eq.~\ref{eq:macc}). Even with high $f_{\rm duty}$, the actual SMBH accretion rate, $\dot{M}_{\rm acc}$, can be small for most of its lifetime, corresponding to a low active fraction. Thus, our choice of higher duty cycle values which set the maximal duty fraction a SMBH can have over its lifetime is compatible with the low observed duty fraction of overall active SMBHs at a given moment. We have checked the accretion rate distribution of our models across redshift, and a majority of SMBHs accrete at $1-0.01$\% of the Eddington rate, consistent with the observed low active fraction. Physically, the duty cycle in our model effectively captures other feedback effects (e.g., mechanical feedback from jets/winds) beyond the thermal feedback considered in our case. Still, our duty cycle parameterization is idealized, assuming a single maximal value for the whole SMBH population throughout cosmic history. Future work will be needed to develop more realistic prescriptions for the AGN duty cycle. Another limitation of our SAM-based modeling is the idealized approach to determining the internal halo gas structure, where we assumed an isothermal profile to estimate the central gas density. Any conclusions here are clearly tentative and need to be further validated with high-resolution cosmological hydrodynamics simulations \citep[e.g.,][]{Sanati2025,Weinberger2025}. Finally, our spectral modeling is exploratory in nature, with quantitative features that depend on uncertain model inputs, to be followed up with a dedicated study in future works as well.


\section{Summary and Conclusions}\label{sec:conclusions}

In this work, we used A-SLOTH, a SAM calibrated to reproduce key high-redshift observations, to model the formation and evolution of a population of SMBHs/AGN in the early Universe, considering heavy DCBH and light stellar-remnant seeds (under multiple modes of accretion-driven growth). In comparing with results from recent {\it JWST} LRD observations, we consider them as a population, as well as focusing on select cases with extreme properties. We find that the LRD population statistics is better matched by the heavy seed models, in particular their observed overmassive nature. Similarly, the extreme properties inferred for CAPERS-LRD-z9 \citep{Taylor2025} and MoM-BH* \citep{Naidu2025} may be more naturally explained by the DCBH formation channel. Sustained super-Eddington accretion on light seeds may also be able to reproduce aspects of the rapidly growing {\it JWST} phenomenology, but this scenario encounters multiple challenges, as discussed above \citep[see also][]{Sacchi2025}.



The emergence of SMBHs in the high-$z$ Universe provides a fundamental stress test for the $\Lambda$CDM model of cosmological structure formation, further accentuating the challenge posed by the abundance of massive, UV-bright galaxies, discovered by {\it JWST} at early cosmic times \citep{MBK2023}. A related key question is whether the bottom-up hierarchy of proceeding from low-mass building blocks can be reversed during the formation of the first SMBHs, as would be the case for heavy DCBH seeds (as well as for some PBH scenarios, e.g., \citealt{Liu2022_PBH,Liu2023,Zhang2025}). This is the broader significance of our proposed scenario, where the Universe can form DCBHs in the most massive and biased halos, thus reproducing select observed LRD population and system properties. We note that multiple distinct scenarios have been proposed to explain LRD sources, other than the pathway discussed here, including but not limited to binary SMBHs \citep{Inayoshi2025_binary}, binaries resulting from PBHs \citep{Zhang2025_PBH_DCBH}, or low-spin halos \citep{Pacucci2025}. They might even not be AGN but rather dense stellar clusters \citep{Leung2024,Guia2024,Perez2024,Baggen2024}, or supermassive stars \citep{Zwick2025}. 

Near-future observations can help to confirm or refute this proposition through metallicity measurements, as we predict that LRDs, if their origins are heavy DCBHs, should be metal-poor while being overmassive at early times \citep[e.g.,][]{Maiolino2025}. New high-redshift metallicity measurements with \textit{JWST} could be forthcoming with updated metallicity calibrations \citep[e.g.][]{Hirschmann2023}. In addition, SMBH populations will be better constrained in the future with gravitational wave observations through pulsar timing arrays \citep{Agazie2023,Antoniadis2023,Reardon2023,Xu2023,Hobbs2017,Romano2017}, and the Laser Interferometer Space Antenna (LISA; \citealt{Robson2019}), which can probe possible massive DCBH seeds at high-$z$. Future X-ray missions, such as Athena \citep{Barret2013} and AXIS \citep{Reynolds2023,Cappelluti2024}, will be able to detect much fainter AGN, compared to existing optical and near-IR surveys, possibly detecting X-ray signatures of LRDs, most of which are currently undetected in X-ray bands \citep{Kocevski2025,Yue2023,Juodbalis2023}.

The remarkable population of LRDs may thus provide the ideal laboratory to disentangle the seeding and growth pathways for the first SMBHs, thus breaking the current degeneracies in physically quite distinct models that can all explain the SMBH phenomenology observed later on in cosmic history \citep[e.g.,][]{Taylor2024}. Upcoming \textit{JWST} surveys promise to unravel the deep mystery of what came first, stars or black holes, as well as elucidating the crucial initial steps of galaxy formation.

\begin{acknowledgments}
The authors acknowledge the Texas Advanced Computing Center (TACC) for providing HPC resources under allocation AST23026. 
BL gratefully acknowledges the funding of the Royal Society University Research Fellowship and the Deutsche Forschungsgemeinschaft (DFG, German Research Foundation) under Germany's Excellence Strategy EXC 2181/1 - 390900948 (the Heidelberg STRUCTURES Excellence Cluster).
\end{acknowledgments}

\appendix

\section{LRD Host Halo Masses}\label{sec:appendix}

In Figure~\ref{fig:mhalo}, we examine how the SMBH host halo mass evolves, showing $M_{\rm halo}$ as a function of redshift.  We further show the evolutionary tracks for the individual cases with the highest $\rho_0$ from Fig.~\ref{fig:rho0gas}. In general, the heavy DCBH seeds reside in more massive halos ($\gtrsim10^7$ M$_\odot$) compared to light stellar remnant seeds, especially at higher redshifts $(z\gtrsim15)$. This is likely due to the virial temperature constraint imposed on DCBH formation (see Section~\ref{sec:seeding}). We show in comparison observational estimates of LRD host halo masses from clustering analysis and abundance matching \citep{Lin2025,Carranza2025}, which align with the more massive and biased halos, hosting massive ($\gtrsim 10^6\ \rm M_\odot$) and luminous ($\gtrsim 10^{44}\ \rm erg\ s^{-1}$) SMBHs, with respect to the general host halo population within our models. Therefore, current LRD observations are likely identifying the most massive and biased host systems.

\bibliography{ms}{}
\bibliographystyle{aasjournal}



\end{document}